\documentstyle[prb,aps,preprint,psfig]{revtex}

\begin{document}
\draft

%@twocolumnfalse\endcsname

\title{Density fluctuations and single-particle dynamics in liquid lithium} 
\author{J. Casas, D. J. Gonz\'alez and L. E. Gonz\'alez} 
\address{Departamento de F\'\i sica Te\'orica, Universidad de Valladolid,
E-47011 Valladolid, SPAIN} 
\author{M.M.G.Alemany and L.J.Gallego } 
\address{Departamento de F\'\i sica de la Materia Condensada, Facultad de 
F\'\i sica, Universidad de Santiago de Compostela, E-15706 Santiago de 
Compostela, SPAIN}
\date{\today}

\maketitle

\begin{abstract}

The single-particle and collective dynamical properties of liquid lithium 
have been evaluated at several thermodynamic 
states near the triple point. This is performed within the framework of 
mode-coupling theory, using a   
self-consistent scheme which, starting from the known 
static structure of the liquid, allows the theoretical 
calculation of several dynamical properties. Special 
attention is devoted to several aspects of the single-particle dynamics,  
which are discused as a function of the thermodynamic state. 
The results are compared with those of 
Molecular Dynamics simulations and other theoretical approaches. 
\end{abstract}


%]

%\narrowtext

\section{Introduction}

For the past twenty years or so, the dynamics of liquid metals has been 
a field of intense research, both theoretical and 
experimental (see, e.g., Ref. \onlinecite{Balubook}),  
especially as regards the liquid alkali metals, all of which
% have been in particular the most thoroughly 
%studied ones; in 
%fact all of them 
have been studied experimentally 
by means of inelastic neutron scattering (INS)
or inelastic X-ray scattering (IXS) or by 
both techniques: Li,\cite{Jong1,Jong2,Seldmeier,BurkelSinn,ScoBal1} 
Na,\cite{MorGla,BalTorStaMor1,Pilg0} 
K,\cite{Novikov} Rb,\cite{CopRow,Pilg1,Pasqualini} and Cs.\cite{MorBo} 
Molecular Dynamics (MD) simulations have also 
stimulated this progress because of their ability to 
determine certain time correlation functions which are not accessible to 
experiment, and thereby they supplement the information obtained from 
the experiments. Most such MD studies have likewise been
performed for alkali  
metals,\cite{npamd,Shimojo,Kahl,BaluTorVall1,TorBalVer} 
although other systems such as  
the liquid alkaline-earths \cite{Alemany} and liquid
lead \cite{Gudowski} have also been studied.

On the theoretical side, this progress can be linked to the development
of microscopic theories which provide a better understanding of the physical
mechanisms underlying the dynamics of simple liquids.\cite{Gotzext} An
important advance in this respect was the observation 
that the decay of several time-dependent
properties can be explained by the interplay of two different dynamical
processes.\cite{Balubook,Sjobook,Sjothself,SjoJPC,Sjothcoll,SjoRb}  
The first one, which gives rise to a rapid initial decay, comprises 
fast, uncorrelated, short-range interactions (collisional
events) which can be broadly identified with ``binary'' collisions. The
second process, which leads to a long-time tail, is connected 
to the non-linear coupling of the dynamical property of interest with slowly
varying collective variables (``modes'') such as  density
fluctuations, currents, etc., and is therefore referred to as a mode-coupling
process. 

By introducing some simplifying approximations, 
Sj\"ogren and coworkers \cite{SjoJPC,Sjothcoll,SjoRb,WahnSjo} first 
applied this 
theory to evaluate several collective and single-particle 
dynamical properties of liquid 
rubidium as well as the single-particle properties of liquid argon, at 
thermodynamic conditions near their respective triple 
points, obtaining results in qualitative agreement with 
the corresponding MD simulations.
%It is also worth mentioning the theoretical studies  
%carried out by 
Balucani and coworkers 
\cite{BaluTorVall1,BaluD} 
also applied a simplified mode-coupling theoretical approach to study
%for 
the liquid alkali metals close 
to their triple points, 
%which basically focused on  
obtaining good qualitative predictions for several 
%some 
transport properties (self-diffusion coefficient and  
shear viscosity) as well as other single-particle 
dynamical properties (velocity autocorrelation function and its memory 
function, and mean square displacement). 
%Their work, which 
%included several simplifications into the mode-coupling theory, showed 
%that the single-particle dynamics of the alkali metals is, at least 
%qualitatively, well described by this theory.
Based on these ideas, we 
developed a theoretical approach \cite{GGCan} which 
allows a self-consistent calculation of all the above transport and 
single-particle dynamical properties; its application to liquid
lithium \cite{GGCan} and the liquid alkaline-earths \cite{Alemany} near
their triple points, has lead to theoretical results in fair agreement
with both simulations and experiment. 

Since single-particle and collective dynamical properties 
are closely interwoven within the mode-coupling theory, the 
application of this formalism to any liquid system should imply 
the self-consistent solution of the coupled equations appearing in 
the theory. However, none of the above-mentioned theoretical 
calculations have 
been performed in this way. In fact, the usual practice   
is to obtain the input dynamical quantities needed for the evaluation of
the mode-coupling expressions either from MD simulations
or from some other theoretical approximation. 
In particular, this has often been the case for the 
intermediate scattering function, 
$F(k, t)$,  which has been obtained either 
from MD simulations,\cite{Shimojo,Gudowski} or
derived from some simple approximation such as the viscoelastic 
model or some simplification of 
it.\cite{BaluTorVall1,TorBalVer,Alemany,BaluD,GGCan}

We have recently presented a theoretical scheme \cite{CaGG} for the  
self-consistent determination of $F(k,t)$ within the mode-coupling theory. 
Its application to study the dynamical properties of liquid 
lithium at thermodynamic states near the triple point 
lead to good qualitative results in comparison  
with MD simulations. However,  the 
self-intermediate scattering function, $F_s(k, t)$, was 
evaluated by means of the gaussian approximation, 
leading to an unsymmetric treatment of $F(k, t)$ and $F_s(k, t)$,  
because mode coupling effects were included for the first but not
for the latter. This imbalance is corrected in the present paper, in which 
we introduce a 
self-consistent framework which treats on an equal footing, and 
within the mode coupling theory, both $F(k, t)$ and $F_s(k, t)$.  
This new scheme is then applied to study the dynamical 
properties of liquid lithium,
%some thermodynamic states close to the triple point. 
covering a somewhat larger range of temperatures than in our previous study.

The only input data required by the theory are the interatomic pair 
potential, which was obtained from the neutral pseudoatom (NPA) 
\cite{npavmhnc} method, and the liquid static structural 
functions, which were evaluated 
through the variational modified hypernetted chain 
approximation (VMHNC) \cite{npavmhnc,Ros86,PRA} theory of liquids. 
Comparison with MD simulations \cite{npamd} has shown that this 
combination leads to an accurate  description of the 
equilibrium properties of liquid lithium close to the triple point. 
We stress that by combining the NPA method to obtain the interatomic  
pair potential, and the VMHNC to calculate the
liquid static structure, we are able to attain the required input data,
for the calculation of the dynamical properties, from the
only knowledge of the atomic number of the system and its thermodynamic
state.

The paper is organized as follows. In section \ref{theory} we describe the
theory used for the calculation of the dynamical properties of the 
system and we propose a self-consistent scheme for the evaluation 
of both single-particle and collective properties.  
In section \ref{results} we 
present the results obtained when this theory is applied to liquid 
lithium at some thermodynamic states. 
Finally we sum up and discuss our results.

\section{Theory}

\label{theory}

This study is based on a combination of kinetic theory ideas with the 
Mori memory function 
approach.\cite{Balubook,Sjobook,Sjothself,SjoJPC,Sjothcoll,SjoRb} 
Within this framework, it is assumed that the  
memory functions of several time correlation functions are  
controlled by the interplay of two different dynamical processes:  
one, with a rapid initial decay, due to
fast, uncorrelated short-range interactions; the  
second (known as a mode-coupling process), 
with a long-time tail, due  
to the non-linear coupling of the specific time correlation function 
with slowly varying collective variables (``modes'') such as 
density fluctuations, currents, etc. 
In the present theoretical framework, we focus on two  
basic dynamical variables, namely, 
the intermediate and the self-intermediate scattering functions which provide 
information about the collective and single-particle 
dynamical properties of the system, respectively. Moreover, their 
respective time-Fourier transforms give the 
dynamic and self-dynamic structure factors which are amenable of 
determination by  means of both INS and IXS experiments.

\subsection{Collective Dynamics}
\label{CollDyn}

The collective dynamical properties are embodied in  
the dynamic structure factor,  
$S(k, \omega)$, 
which can be obtained as 

\begin{equation}
S(k, \omega) = \frac{1}{\pi} \; Re \; \tilde{F}(k, z= -i\omega) \, ,
\end{equation}

\noindent where $Re$ stands for the real part and 
$\tilde{F}(k, z)$ is the Laplace transform of the  
intermediate scattering function, $F(k, t)$, i.e., 

\begin{equation}
\label{LapF}
\tilde{F}(k, z)= \int_0^{\infty} \; dt \; e^{-zt} F(k, t) \, .
\end{equation}

\noindent Within the memory function formalism, $\tilde{F}(k,z)$ 
can be expressed as \cite{SjoJPC}
 
\begin{equation}
\label{MfF}
\tilde{F}(k, z) = S(k)\left[ z + \frac{\Omega^2(k)}{z 
+\tilde{\Gamma}(k, z)}\right]^{-1} \, ,
\end{equation}

\noindent where $\tilde{\Gamma}(k, z)$ is the Laplace transform of the 
second-order memory function, $\Gamma(k, t)$, and 
%\begin{equation}
$\Omega^2(k)=k^2/\beta m S(k)  \, $
%\end{equation}
%\noindent 
where $m$ is the mass of the particles, $\beta$ is the inverse 
temperature times the Boltzmann constant and $S(k)$ is the 
static structure factor of the liquid. Note that since we 
consider spherically symmetric potentials and homogenous systems, 
the dynamical magnitudes only depend on the 
modulus $ k = \mid \vec{k} \mid $.

Now, the second-order memory function $\Gamma(k, t)$ is decomposed  
as follows:\cite{Balubook,Sjothcoll,SjoRb}  
 
\begin{equation}
\label{Gammatot}
\Gamma(k, t) = \Gamma_{B}(k, t) + \Gamma_{\rm MC}(k, t) \, ,
\end{equation}
 
\noindent where 
the term $\Gamma_{B}(k, t)$, known as the binary part, has a 
fast decay,
whereas the the mode-coupling 
contribution, $\Gamma_{\rm MC}(k, t)$, which aims to take into account 
repeated correlated 
collisions, starts as $t^4$, reaches a maximum and then decays rather 
slowly. We now briefly describe both terms; for more details we 
refer the reader to Ref. 
\onlinecite{Balubook,Sjobook,Sjothself,SjoJPC,Sjothcoll,SjoRb}.

\subsubsection{The binary-collision term} 
Over very short times, the 
memory function is well described by 
$\Gamma_{B}(k, t)$ alone, both 
$\Gamma(k, t)$ and $\Gamma_{B}(k, t)$ having the same initial value and
curvature:\cite{Balubook}

\begin{equation}
\label{Gambin}
\Gamma(k,0)= \Gamma_{B}(k,0)= \frac{3 k^2}{\beta m} + \Omega_0^2 + 
\gamma_d^l(k) - \Omega^2(k) \, ,
\end{equation}

\noindent where  

\begin{equation}
\Omega_0^2 = \frac{\rho}{3 m} \int d\vec{r} \, \, g(r)\, 
{\nabla}^2 \varphi (r) 
\end{equation}

\noindent is the squared Einstein frequency and 

\begin{equation}
\gamma_d^l(k) = - \frac{\rho}{m} \int d\vec{r} \, 
e^{-i \vec{k} \cdot \vec{r}} 
%g(r) \, \frac{\partial^2}{\partial z^2} \varphi (r)
g(r) \, (\hat{k} \cdot \vec{\nabla})^2 \varphi(r) \, ,
\end{equation}

\noindent with $\varphi (r)$ and $g(r)$ denoting respectively the 
interatomic pair potential and the pair distribution function of the liquid 
system with number density $\rho$, and $\hat{k}=\vec{k}/k$.

The binary term includes all the contributions to $\Gamma(k, t)$ to 
order $t^2$. Since 
the detailed features of the ``binary'' dynamics of systems with continuous
interatomic potentials are rather poorly known, we resort 
to a semi-phenomenological
approximation that reproduces the correct 
short-time expansion. Therefore we write \cite{CaGG}

\begin{equation}
\label{gammaB}
\Gamma_{B}(k, t)= \Gamma_{B}(k,0)  
\; e^{-t^2/\tau_l^2(k)} \, ,
\end{equation}

\noindent where the relaxation time, $\tau_l(k)$, 
%is a relaxation time, whose form 
can be determined from a short-time expansion of the formally 
exact expression of the binary term, and 
%the result 
is related to 
the exact sixth frequency moment of $S(k, \omega)$. In this way, 
after making the superposition approximation for the three-particle 
distribution function, one obtains 
%\cite{Sjothcoll,SjoRb,Balubook}
\cite{Balubook}

\begin{eqnarray}
\label{Tauele}
& & \frac{\Gamma_{B}(k,0)}{\tau_l^2(k)}  = \nonumber \\
& & \frac{3 k^2}{2 \beta m} 
\left[ \frac{2 k^2}{\beta m} 
+ 3 \Omega_0^2 +  
2 \gamma_d^l(k)\right] + \nonumber \\
& & (\frac{3 \rho}{\beta m^2})\, i k \int d \vec{r}\, 
e^{-i \vec{k} \cdot  \vec{r}} 
g(r)(\hat{k} \cdot \vec{\nabla})^3 \varphi(r) +
\nonumber \\  
& & \frac{\rho}{m^2} \int d \vec{r}\, [1 - e^{-i\vec{k}\cdot\vec{r}}] \,
[{\hat{k}}^{\alpha}\nabla^{\alpha}\nabla^{\gamma}\varphi(r)]\, g(r) \,
[{\hat{k}}^{\beta}\nabla^{\beta}\nabla^{\gamma}\varphi(r)] + \nonumber \\ 
& & \frac{1}{2 \rho} \int \frac{d {\vec{k}}^{\prime}}{(2 \pi)^3} 
{\hat k}^{\alpha} \gamma_d^{\alpha \gamma}({k}^{ \prime}) 
\left\{ \left[ S({k}^{ \prime}) -1 \right] + \left[S(\mid \vec{k}  - 
\vec{{k}^{ \prime}}\mid )-1 \right] \right\} 
\nonumber \\
& & \times \left\{ \gamma_d^{\beta \gamma}({k}^{ \prime})-
\gamma_d^{\beta \gamma}
(\mid \vec{k}-\vec{{k}^{ \prime}}\mid ) \right\} {\hat{k}}^{\beta} \, ,
\nonumber \\
& & 
\end{eqnarray}

\noindent where %$S(k)$ is the liquid static structure factor and 
summation over repeated indices 
is implied ($\alpha, \beta, \gamma = x, y, z$), and  

\begin{equation}
\gamma_d^{\alpha \beta}(k) = - \frac{\rho}{m} \int d \vec{r} \,\, 
e^{-i \vec{k} \cdot \vec{r}}  \, g(r) \, \nabla^{\alpha} \nabla^{\beta} 
\varphi (r) \, .
\end{equation}

%\noindent and therefore, 
The relaxation time can thus be evaluated from the 
knowledge of the interatomic pair potential and its 
derivatives together with the 
static structural functions of the liquid system.

\subsubsection{The mode-coupling component} 
The inclusion of a slowly decaying time tail is essential
in order to achieve, at least, a qualitative description of $\Gamma(k, t)$. 
A detailed treatment of this term requires consideration of   
several modes (density-density coupling, density-longitudinal 
current coupling and density-transversal current coupling). However, 
for thermodynamic conditions 
near the triple point, the most important contribution
is the density-density term, which can be 
written as \cite{SjoRb}

\begin{eqnarray}
%\nonumber
\label{gammamc}
& &  \Gamma_{\rm MC}(k, t)  = %\nonumber \\
% & &  
\frac{\rho}{\beta m} \int \frac{d\vec{k^{\prime}}}
{(2 \pi)^3}\, \hat{k} \cdot \vec{k^{\prime}} \, c(k^{\prime}) \nonumber \\
& & \times
\left[ \hat{k} \cdot \vec{k^{\prime}} \, c(k^{\prime}) + 
\hat{k} \cdot (\vec{k}-\vec{k^{\prime}}) \, 
c(\mid k-k^{\prime} \mid)\right] \nonumber \\ 
& & \times \left[F( \mid \vec{k} - \vec{k^{\prime}} \mid, t) \, 
F(k^{\prime}, t) - 
F_B(\mid \vec{k} - \vec{k^{\prime}} \mid, t) \, F_B(k^{\prime}, t)\right]
\, ,
\nonumber \\
& & 
\end{eqnarray}

\noindent where $c(k)$ is the direct correlation function and   
$F_B(k, t)$ denotes the binary part of 
the intermediate scattering function, $F (k, t )$. 
Following Sj\"ogren,\cite{SjoRb} we approximate the ratio 
between $F(k, t)$ and 
its binary part by the ratio between their corresponding 
self counterparts, i.e.,  

\begin{equation}
\label{FB}
F_B(k, t) =\frac{F_{s,B}(k, t)}{F_s(k, t)} \,  
F(k, t) \, ,
\end{equation}

\noindent where $F_s(k, t)$ is the self-intermediate scattering function 
and $F_{s,B}(k, t)$ stands for its binary part, which is now approximated 
by the free-particle expression   

\begin{equation}
F_{s,B}(k, t) = F_0(k, t) \equiv \exp [-\frac{1}{2m \beta}k^2t^2] \, . 
\end{equation}

Once $F_s(k, t)$ has been specified, the self-consistent evaluation 
of the above formalism will yield to $F(k,t)$, and, therefore 
to all the collective dynamical properties discussed 
so far.

\subsubsection{Other models}

%It is worth mentioning another, much simpler and widely used model for 
%the second-order memory function, $\Gamma(k, t)$; this is the so-called 
%viscoelastic model. It approximates $\Gamma(k, t)$ by an 
%exponentially decaying function with a single relaxation time, which is 
%usually fitted so that the predicted value of $S(k, \omega=0)$ coincides for 
%$k \to \infty$ with the exact free-particle 
%result.\cite{Copley&Lovesey,Lovesey} This model  
%has been used in previous studies of single-particle dynamics, within
%the mode-coupling formalism, as a simple expression 
%for the intermediate scattering function. 
%In this work, we will also use the viscoelastic model, with the relaxation 
%time proposed by Lovesey,\cite{Lovesey} in order to compare its predictions 
%with those derived from the more elaborated self-consistent formalism
%proposed above. 

In this work, we will compare the predictions of the above self-consistent 
approach with those of a 
much simpler and widely used model for 
the second-order memory function, $\Gamma(k, t)$; this is the so-called 
viscoelastic model. It approximates $\Gamma(k, t)$ by an
exponentially decaying function with a single relaxation time, which is 
usually fitted so that the predicted value of $S(k, \omega=0)$ coincides for 
$k \to \infty$ with the exact free-particle 
result.\cite{Copley&Lovesey,Lovesey}

\subsection{Single-particle dynamics}

\label{single}

The self-intermediate scattering function, $F_s(k, t)$,
probes the single-particle dynamics 
over different length scales, ranging from the hydrodynamic limit 
($k \to 0$) to the free-particle limit ($k \to \infty$). 
Its frequency spectrum is the self-dynamic structure 
factor, $S_s(k, \omega) = (1/\pi) \; Re \; \tilde{F}_s(k, z= -i \omega)$,
where 
$\tilde{F}_s(k, z)$ stands for the Laplace transform, which according to 
the memory function formalism, can be expressed as

\begin{equation}
\label{MfFs}
\tilde{F}_s(k, z) = 
\left[ z + \tilde{K}_s(k, z) \right]^{-1} \,=  
\left[ z + \frac{\Omega_s^2(k)}{z +\tilde{\Gamma}_s(k, z)}\right]^{-1} \, ,
\end{equation}

\noindent where 
$\Omega_s^2(k)=k^2/(\beta m)$ and   
$\tilde{K}_s(k,z)$ and 
$\tilde{\Gamma}_s(k, z)$ are respectively the Laplace transforms 
of the first- and  
second-order memory functions of $F_s(k, t)$.  
%At this stage, it is worth mentioning that 
Note that
the velocity autocorrelation 
function (VACF) 
of a tagged particle in the fluid, 
i.e., $\frac{1}{3} \; 
\langle \vec{v}_1(t) \vec{v}_1(0) \rangle $, 
is also given by the limit 
$\lim_{k \to 0} 
K_s(k, t)/k^2$.  
As before, 
$\Gamma_s(k, t)$ is now decomposed\cite{Balubook,Sjothcoll,SjoRb}  
 
\begin{equation}
\label{Gammastot}
\Gamma_s(k, t) = \Gamma_{s,B}(k, t) + \Gamma_{s,\rm MC}(k, t) \, ,
\end{equation}
 
\noindent into a binary contribution, 
$\Gamma_{s,B}(k, t)$ and  a mode-coupling term,
$\Gamma_{s, \rm MC}(k, t)$.

\subsubsection{The binary-collision term}

The binary term shows a fast decay, 
$\Gamma_s(k, t)$ and $\Gamma_{s,B}(k, t)$ having the same initial value and
curvature:\cite{Balubook}

\begin{equation}
\Gamma_{s }(k, 0)= \Gamma_{s,B}(k,0) =  
\frac{2 k^2}{\beta m} + \Omega_0^2 \,\,\,\, . 
\end{equation}

Its time dependence can be adequately described by 
a semi-phenomenological expression, 

\begin{equation}
\label{gammasB}
\Gamma_{s, B}(k, t)= \Gamma_{s,B}(k,0)  
\; e^{-t^2/\tau_s^2(k)} \, ,
\end{equation}

\noindent where $\tau_s(k)$ is a relaxation time, 
determined from a short time expansion of the formally 
exact expression of the binary term, which can be related to the
sixth frequency moment of $S_s(k, \omega)$. In this way, 
after making the superposition approximation for the three-particle 
distribution function, one obtains
%\cite{Sjothcoll,SjoRb,Balubook}
\cite{Balubook}

\begin{equation}
\label{Tauese}
 \frac{\Gamma_{s,B}(k,0)}{\tau_s^2(k)}  = 
 \frac{3 k^2}{2 \beta m} 
\left[ \frac{2 k^2}{\beta m} 
+ 3 \Omega_0^2 
\right] + \frac{ \Omega_0^2 }{\tau^2}
\end{equation}

\noindent where

\begin{eqnarray}
\label{Tau}
\frac{\Omega_0^2}{{\tau}^2} & = &
\frac{\rho}{3 m^2} \int d \vec{r}\, 
[\nabla^{\alpha}\nabla^{\beta}\varphi(r)]\, g(r) \,
[\nabla^{\alpha}\nabla^{\beta}\varphi(r)]  \nonumber \\
& + & \frac{1}{6 \rho} \int \frac{d {\vec{k}}^{\prime}}{(2 \pi)^3} 
\gamma_d^{\alpha \beta}(k^{ \prime}) 
\ [ S(k^{ \prime}) -1 ]  
 \gamma_d^{\alpha \beta }(k^{ \prime}) \,\,\,\, , 
\end{eqnarray}
  
\noindent with summation over repeated indices implied.

\subsubsection{The mode-coupling component}

This term describes the intermediate and long time behaviour, involving 
repeated correlated collisions. 
%Although several 
%modes could be included, we follow the same recipe as with 
As for its collective 
counterpart, we simplify the full multi-mode expression by considering only the 
coupling to the density fluctuations:\cite{Balubook}

\begin{eqnarray}
%\nonumber
\label{gammasmc}
& &  \Gamma_{s,\rm MC}(k, t)  = %\nonumber \\
% & &  
\frac{\rho}{\beta m} \int \frac{d\vec{k^{\prime}}}
{(2 \pi)^3}\, (\hat{k} \cdot \vec{k^{\prime}})^2 \, c^2(k^{\prime}) \nonumber \\
& & \times \left[F_s( \mid \vec{k} - \vec{k^{\prime}} \mid, t) \, 
F(k^{\prime}, t) - 
F_{s,B}(\mid \vec{k} - \vec{k^{\prime}} \mid, t) \, F_B(k^{\prime}, t)\right]
\, .
\nonumber \\
& & 
\end{eqnarray}

%\noindent We note that the 
%part involving the product of 
%$F_{s, B}(\mid \vec{k} - \vec{k^{\prime}} \mid, t )$ $F_B(k^{\prime}, t)$ has 
%the effect of making 
%$\Gamma_{s,\rm MC}(k, t)$ very small at short times; therefore the influence of the 
%approximation made for $F_B(k, t)$ is negligible beyond a short time 
%interval, because the intermediate and long time features of 
%$\Gamma_{s,\rm MC}(k, t)$ are controlled by the term involving 
%the product of   
%$F_{s}(\mid \vec{k} - \vec{k^{\prime}} \mid, t )$ $F(k^{\prime}, t)$.  

\noindent Note that the term involving the product $F_{s,B} F_B$, which
makes $\Gamma_{s,\rm MC}(k, t)$ very small at short times, decays very
fast. Therefore the possible errors introduced by making specific
approximations for these binary terms become negligible in the region of
intermediate and long times, where the dominant contribution to
$\Gamma_{s,\rm MC}(k, t)$ is the product $F_s F$.

\subsubsection{Other models}

A viscoelastic model has also been proposed for the 
$F_s(k, t)$; it assumes for $\Gamma_s(k, t)$ an 
exponentially decaying function with a single relaxation 
time, which can in principle be fixed using several prescriptions
that have been proposed.
In this work we will use the expression obtained bu requiring that
$S_s(k, \omega=0)$ coincides for $k \to \infty$ with the exact 
free-particle result.\cite{Loveseyself} 

A different, more accurate and widely used, model 
for $F_s(k, t)$ is the so-called gaussian approximation (GA):

\begin{equation}
\label{Fsgauss}
F_s(k, t)= \exp \left[ \; -   \frac{1}{6} \; k^2 \; \delta r^2(t) \right] 
%F_s(k, t)= exp \left[ -  \left( \frac{k^2}{m \beta} \right) 
%\int_0^t \; d\tau \; ( t- \tau) Z(\tau) \right]  \,\,\,\, ,
\end{equation}

\noindent where $\delta r^2(t) \equiv 
\langle | \vec{r}_1(t) - \vec{r}_1(0) |^2 \rangle$ stands for 
the mean squared displacement of a tagged particle in the fluid. This 
approximation produces correct results in the limits of both small and large 
wavevectors, and for all wavevectors at short times. 
Moreover, $\delta r^2(t)$ is 
related to the normalized VACF, 
$Z (t) = \langle \vec{v}_1(t) \vec{v}_1(0) \rangle
/ \langle v_1^2 \rangle $, by                                                   

\begin{equation}
\label{r2t}
\delta r^2(t) = \frac{6}{\beta m} \int_0^t d \tau \; (t - \tau) \; Z(\tau)
\,\,\,\, .
\end{equation}

%\noindent In fact, in a previous work,\cite{CaGG} we presented a 
%self-consistent 
%theoretical scheme which combined the above expressions for  
%$F(k, t)$  (see Eq. (\ref{MfF})), along with the 
%GA model of Eq. (\ref{Fsgauss}) for  $F_s(k, t)$. 
%The evaluation of $\delta r^2(t)$  in 
%Eq. (\ref{Fsgauss}) was performed in terms 
%of $Z(t)$ (see Eq. (\ref{r2t})), by  
%resorting to its first order memory function, which is a 
%time-dependent function obtained as the $k \to 0$ limit of the present 
%$\Gamma_s(k, t)$ (see Eq. (\ref{Gammastot})). That combination resulted 
%in an unsymmetric treatment of the intermediate scattering functions, since  
%the mode-coupling effects were explicitely included for $F(k, t)$  but 
%not for  $F_s(k, t)$.   

\subsection{Self-consistent procedure}

\label{self}

The calculations begin with some estimation for both $F(k, t)$ and $F_s(k, t)$ 
(e.g., those estimates provided by the corresponding viscoelastic models). 
Using the known values of 
$\Gamma_{s,B}(k, t)$ and Eq. (\ref{gammasmc}) and (\ref{FB}), a 
total memory function $\Gamma_s(k, t)$ is obtained which, when taken to 
Eq. (\ref{MfFs}), gives a new estimate for $F_s(k, t)$. This is now 
taken to Eq. (\ref{gammamc}) which, along with the 
known values of $\Gamma_{B}(k, t)$, produces an estimate for 
the total memory function $\Gamma(k, t)$, which is taken to 
Eq. (\ref{MfF}) for a new determination of $F(k, t)$. Now, with 
the new estimates for both  
$F(k, t)$ and $F_s(k, t)$  the whole procedure is repeated.

The previous computational loop is iterated until self-consistency 
is achieved between the initial
and final $F(k, t)$ and $F_s(k, t)$. The practical application of this scheme 
has shown that reaching self-consistency requires about ten iterations.  

\subsection{Transport properties}
\label{shear}

Once the self-consistency has been achieved, the normalized 
VACF is obtained 
as 

\begin{equation}
\label{VACF}
Z(t)= (\beta m) \; \lim_{k \to 0} \frac{K_s(k, t)}{k^2}
\end{equation}

\noindent whereas its associated transport coefficient, namely the 
self-diffusion coeficient, $D$, is given by 

\begin{equation}
\label{DifZ}
D = \frac{1}{\beta m} \int_0^{\infty} dt \, Z(t) \, .
\end{equation}

%In this paper, we apply this scheme to study liquid lithium at several 
%thermodynamic states as well as the liquid alkaline-earth metals near their 
%correponding triple points. The results obtained within this scheme will 
%usually be compared with those predicted by the viscoelastic approximation, 
%that is, those results obtained using the viscoelastic $F(k, t)$ and 
%$F_s(k, t)$.  

%\bigskip

%\subsection{Shear viscosity}

Another interesting transport property is the shear viscosity 
coefficient, $\eta$, which can be obtained as the time integral of the 
stress autocorrelation function (SACF), $\eta(t)$, which
stands for the time autocorrelation function of the non-diagonal elements 
of the stress tensor. 
Moreover, $\eta(t)$ can be decomposed into three contributions, a
purely kinetic term, $\eta_{kk}(t)$, a purely potential 
term, $\eta_{pp}(t)$, and a crossed term, $\eta_{kp}(t)$. However,  
for the liquid range close to the triple point,  
the contributions to $\eta$ coming from the first and last
terms are negligible,\cite{Balueta} and therefore in the present 
calculations we assume
$\eta(t) = \eta_{pp}(t)$.
%, the purely
%potential part of the SACF. 
This function can in turn be split
into its binary and mode-coupling
components, 
$\eta(t) = \eta_B(t) + \eta_{\rm MC}(t)$. Again, the binary part 
is described by means of a gaussian ansatz, i.e.,   

\begin{equation}
\eta_B(t) = G_p \; e^ {-t^2/\tau_{\eta}^2} \, , 
\end{equation}

\noindent where the rigidity modulus, $G_p$, and 
%is the initial value of both  
%$\eta(t)$ and $\eta_B(t)$, and 
the initial time decay, $\tau_{\eta}$, %is their initial time 
%decay. As shown in Ref.\  \onlinecite{GGCan}, both $G_p$ and $\tau_{\eta}$  
can both be computed from the 
%knowledge of the 
interatomic potential and 
the static structural functions of the system.\cite{GGCan} 
The superposition approximation 
for the three-particle distribution function is also used in the evaluation 
of $\tau_{\eta}$.   
For the mode-coupling component, $\eta_{\rm MC}(t)$, we only consider 
the coupling to density fluctuations, 
given by,\cite{Balubook,GGCan,Balueta}

\begin{eqnarray}
& &\eta_{\rm MC}(t) = \nonumber \\
& &\frac{1}{60 \beta \pi^2} \int dk \, k^4 \left[ \frac{S
^{\prime}(k)}{S^2(k)} \right]^2 \left[ F^2 (k,t) - 
F_B^2(k,t) \right] \, ,
\end{eqnarray}

\noindent where %$F_B(k, t)$ is given by Eq. (\ref{FB}) and  
$S^{\prime}(k)$ is the derivative of the static 
structure factor with respect to $k$. This  
mode-coupling integral is evaluated using the $F(k, t)$ and $F_B(k, t)$ 
previously obtained within the self-consistent scheme described above.

\section{Results}

\label{results}

We have applied the preceeding theoretical formalism to study the dynamical
properties of liquid $^7$Li at several thermodynamic states near  
the triple point (see Table I).
The input data required for the calculation of the 
dynamical properties are 
both the interatomic pair potential and its derivatives, 
which were evaluated by the the NPA \cite{npavmhnc} method, 
as well as the liquid static structural properties, 
evaluated by the VMHNC \cite{Ros86,PRA} theory of liquids.  
%%********************************************
%\begin{figure}
%\begin{center}
%\mbox{\psfig{file=f1.eps,angle=-90,width=75mm}}
%\end{center}
%\caption{Relaxation times $\tau_l(k)$ and $\tau_s(k)$,  
%for liquid lithium. 
%Continuous line and dotted line:  $\tau_l(k)$ and $\tau_s(k)$ for T= 470 K. 
%Dashed line and dash-dotted line:  $\tau_l(k)$ and $\tau_s(k)$ for T= 725 K.} 
%\label{Taus}
%\end{figure}
%%********************************************

First, we have evaluated the relaxation times appearing in the binary part 
of the second order memory functions of the intermediate and self-intermediate 
scattering functions, $\tau_l(k)$ and $\tau_s(k)$, respectively,  
using Eq. (\ref{Tauele}) and 
(\ref{Tauese}). The obtained results are plotted in Fig. \ref{Taus}, which
shows that for each temperature, both relaxation times have
approximately the same magnitude, with  $\tau_l(k)$ 
oscillating around $\tau_s(k)$.

\subsection{Intermediate scattering function} 

Figures \ref{MEMFKT470} and \ref{MEMFKT725} show, 
for $T$=470 K and 725  K,  
at two $k$ values, 
the %MD results obtained for the 
first and second-order memory functions of $F(k, t)$, as obtained from 
the MD simulations,  
and the corresponding theoretical ones derived  
within the present approach. 
%%*************************************************************
%\begin{figure}
%\begin{center}
%\mbox{\psfig{file=f2.eps,angle=-90,width=65mm}}
%\end{center}
%\caption{Normalized second-order memory function, $\Gamma (k, t)$, of 
%the intermediate scattering function, $F(k, t)$, at  
%two $k$-values, for liquid lithium at T = 470 K. Open circles: MD results. 
%Continuous line: present theory. 
%Dash-dotted line: binary part, $\Gamma_B(k, t)$. 
%Dotted line: mode-coupling part, $\Gamma_{\rm MC}(k, t)$. 
%Dashed line: viscoelastic model. The inset shows the normalized 
%first-order memory 
%function as obtained by MD (open circles), the viscoelastic 
%model (dashed line) and the present theory (continuous line).}
%\label{MEMFKT470}
%\end{figure}
%%**************************************************************
The MD results for the second-order memory function, $\Gamma_{\rm MD}(k, t)$,  
show a rapid initial decay followed by 
a long-time tail which becomes 
%longer for the smaller $k$-values, 
%(ii) it exhibits an oscillatory behaviour for $k \geq  k_p/2$ and 
%(iii) its relevance diminishes for increasing temperatures. 
smaller when the wavevector or the temperature increases, and
exhibits an oscillatory behaviour for $k \geq  k_p/2$.
The MD results for the first-order memory function show 
a negative minimum whose absolute value decreases as $k$ or 
$T$ increases. 
%%***********************************************************
%\begin{figure}
%\begin{center}
%\mbox{\psfig{file=f3.eps,angle=-90,width=65mm}}
%\end{center}
%\caption{Same as the previous figure but for T = 725 K. } 
%\label{MEMFKT725}
%\end{figure}
%%***********************************************************
The theoretical $\Gamma (k, t)$ obtained by the self-consistent 
procedure qualitatively reproduce the 
corresponding MD results, especially the short time behaviour, which 
is dominated by the binary component. For larger $t$ the   
overall amplitude of the decaying 
tail of $\Gamma(k,t)$ is, in general, well described, but
some discrepancies with the MD results appear for the amplitudes of
the oscillations. We believe that its improvements will 
require the inclusion of other terms in the 
expression for $\Gamma_{\rm MC}(k, t)$ (see Eq. (\ref{gammamc})),
in particular those related to the density-currents couplings. 
The viscoelastic model,\cite{Copley&Lovesey}  
whose results, $\Gamma_{\rm visc}(k, t)$, are  also shown
in Figs. \ref{MEMFKT470} and \ref{MEMFKT725},
%Here, this 
%function is described by a simple exponential decaying function which is 
is clearly unable to describe the features exhibited by 
$\Gamma_{\rm MD}(k, t)$.   

%%**************************************************
%\begin{figure}
%\begin{center}
%\mbox{\psfig{file=f4.eps,angle=-90,width=85mm}}
%\end{center}
%\caption{Normalized intermediate scattering functions, $F(k, t)$, at  
%several $k$-values, for liquid 
%lithium at T = 470 K. Open circles: MD results. 
%Continuous line: present theory. 
%Dashed line: viscoelastic model.}
%\label{FKT470}
%\end{figure}
%%************************************************************

The corresponding MD results for the 
intermediate scattering functions, $F_{\rm MD}(k, t)$, 
are shown in Figs. \ref{FKT470} and \ref{FKT725}. 
Below $k \approx k_p/2$, 
$F_{\rm MD}(k, t)$ exhibit an oscillatory behaviour, with the amplitude 
of the oscillations being stronger for the smaller $k$-values. 
These features are predicted quite well by the  
self-consistent theoretical $F(k, t)$, 
especially as regards the amplitude of the decaying tail, which is  well
reproduced. The main discrepancy %are restricted to the small 
%$k$-values where the amplitude of the oscillations is underestimated. 
is the underestimation of the amplitude of the oscillations for small $k$.
These results, however, represent an important improvement over 
those obtained from the viscoelastic model,\cite{Copley&Lovesey}
$F_{\rm visc}(k, t)$, which 
basically oscillate around zero. 
Moreover, for small $k$-values, i.e., $k \approx 0.25$ \AA$^{-1}$, 
the oscillations in  $F_{\rm visc}(k, t)$ are overdamped, whereas
for the intermediate $k$-values, 
i.e., $k \approx 1$ \AA$^{-1}$ and $k \approx 1.7$ \AA$^{-1}$, they
are too strong.
Only for $k$-values around $k_p$ the viscoelastic results  
show a good agreement with the MD ones. 
This agreement 
%for the regions around the main peak 
can be understood
as a compensation between the short-time and intermediate-time
deficiencies of  $\Gamma_{\rm visc}(k, t)$. 
On the other hand, note that the viscoelastic 
model imposes the exact initial values of $F(k, t)$, and its 
second and fourth derivatives; this explains the good agreement obtained at 
short times. 

%%****************************************************
%\begin{figure}
%\begin{center}
%\mbox{\psfig{file=f5.eps,angle=-90,width=85mm}}
%\end{center}
%\caption{Same as the previous figure but for T = 725 K. } 
%\label{FKT725}
%\end{figure}
%%******************************************************************

Finally, we point out that the theoretical results for 
$F(k, t)$ and its corresponding memory functions, are rather similar to the 
those %results presented in a previous work,\cite{CaGG}  where 
% $F(k,t)$ was described by the same formalism of section 
%\ref{CollDyn}, whereas 
obtained using the GA model for  $F_s(k, t).$\cite{CaGG}
This is not unexpected, because the role of  $F_s(k, t)$ in the determination 
of  $\Gamma (k, t)$ is restricted to the evaluation of $F_B(k, t)$ 
appearing in  $\Gamma_{\rm MC}(k, t)$ in 
Eq. (\ref{gammamc}). However, 
the part involving the product of the $F_B(k, t)$'s just 
makes $\Gamma_{\rm MC}(k, t)$ very small at short times, where, in any
case, the dominant contribution is the binary term, $\Gamma_{B}(k, t)$. 
Therefore the possible errors in $F_s(k,t)$ due to the GA are of little 
account.

%\bigskip

%\bigskip

\subsection{Self intermediate scattering function}

Figures \ref{MEMFsKT470} and \ref{MEMFsKT725} show, 
for T=470 and 725  K,  the  MD and theoretical 
results obtained for the first and second-order memory functions  
of $F_s(k, t)$.

%%****************************************************
%\begin{figure}
%\begin{center}
%\mbox{\psfig{file=f6.eps,angle=-90,width=85mm}}
%\end{center}
%\caption{Normalized second-order memory function, $\Gamma_s (k, t)$, of 
%the self intermediate scattering functions, $F_s(k, t)$, at  
%several $k$-values, for liquid lithium at T = 470 K. Open circles: MD results. 
%Continuous line: present theory. 
%Dotted line: mode-coupling part, $\Gamma_{s, \rm MC}(k, t)$. 
%Dashed line: viscoelastic model. 
%Dash-dotted line: GA model for  $F_s(k, t)$.  
%The inset shows the normalized  
%first-order memory 
%function as obtained by MD (open circles), the viscoelastic 
%model (dashed line) and the present theory (continuous line).}
%\label{MEMFsKT470}
%\end{figure}
%%*******************************************************************

The MD results for the second-order memory 
function, $\Gamma_{s, \rm MD}(k, t)$, are qualitatively
very similar to those previously
obtained for  $\Gamma_{\rm MD}(k, t)$: it exhibits a rapid decaying part 
at short times, followed by a long-time tail which can take negative values for 
the greater $k$-values. However, the long-time tail of 
$\Gamma_{s, \rm MD}(k,t)$ oscillates 
for all $k$-values, not just for $k \geq k_p/2$, as in the case of 
$\Gamma_{\rm MD}(k,t)$.
Also, we note that the role of the long-time tail becomes less relevant 
when the temperature is increased. 
The first-order memory function has
%of $F_s(k, t)$, we observe the appearance of 
a negative minimum for all the $k$-values, followed by an oscillating tail. 
In fact, similar qualitative features were 
also obtained by Shimojo {\it et al.} \cite{Shimojo} 
in their MD study for liquid Na near the triple point.  
The self-consistently calculated $\Gamma_s(k, t)$ 
shows a good 
overall agreement with the MD results. 
The short-time behaviour, which is dominated by the binary 
component, is well reproduced, since the present theoretical approach 
imposes the exact 
initial values of both $\Gamma_s(k,t)$ and its second derivative. 
The intermediate and long-time behaviour, which is controlled by 
$\Gamma_{s, \rm MC}(k, t)$, is qualitatively well described, especially 
the oscillations of the tail for small $k$-values. However, 
increasing the temperature (Fig. \ref{MEMFsKT725}) leads to
an overestimation of $\Gamma_s(k,t)$ for intermediate and long times.
This fact may signal the need to incorporate other 
coupling terms in the expression of $\Gamma_{s, \rm MC}(k, t)$. 
%%******************************************************
%\begin{figure}
%\begin{center}
%\mbox{\psfig{file=f7.eps,angle=-90,width=85mm}}
%\end{center}
%\caption{Same as the previous figure but for T = 725 K. } 
%\label{MEMFsKT725}
%\end{figure}
%%**************************************************************
For comparison, we have also included in 
Figs. \ref{MEMFsKT470} and \ref{MEMFsKT725}  
the results of the viscoelastic model \cite{Loveseyself} for 
the second-order memory function, $\Gamma_{s, \rm visc}(k, t)$. It is 
observed that 
this function provides a rather poor description. In fact, for $k \leq k_p$ 
$\Gamma_{s, \rm visc}(k, t)$ underestimates the MD results for all times, 
whereas for larger 
$k$-values $\Gamma_{s,\rm visc}$ is too small for short times, and too
large at the intermediate times.  
A further comparison is also performed with 
the GA model, as given by Eq. (\ref{Fsgauss}), using 
the $\delta r^2(t)$ obtained from the MD simulations. We have calculated the 
corresponding second-order memory functions, $\Gamma_{s, g}(k, t)$, 
which are also shown in 
Figs. \ref{MEMFsKT470} and \ref{MEMFsKT725}. It is observed that 
$\Gamma_{s, g}(k, t)$ gives a reasonable account of the 
corresponding MD results, especially for short and intermediate times,
whereas for longer times the model underestimates the MD results. 

%%****************************************
%\begin{figure}
%\begin{center}
%\mbox{\psfig{file=f8.eps,angle=-90,width=85mm}}
%\end{center}
%\caption{Self intermediate scattering functions, $F_s(k, t)$, at  
%several $k$-values, for liquid lithium at T = 470 K. Open circles: MD results. 
%Continuous line: present theory. Dashed line: viscoelastic model. 
%Dash-dot line: GA model }
%\label{FsKT470}
%\end{figure}
%%*******************************************************

The results obtained for the self-intermediate scattering function 
are shown in Figs. \ref{FsKT470} and \ref{FsKT725} for 
several $k$-values and 
T=470 and 725 K. The MD results, $F_{s, \rm MD}(k, t)$, decreases 
monotonically with time for all the $k$-values, and this 
behaviour is rather well described by the present theoretical formalism, 
which leads to $F_s(k, t)$ which closely follow the corresponding MD results. 
On the other hand, the viscoelastic model leads to 
$F_{s, \rm visc}(k,t)$ which underestimate the MD results seriously, 
especially for the smaller $k$-values. This can be explained because the 
viscoelastic model incorporates the 
exact initial values of $F_s(k, t)$, and its second and fourth derivatives, 
but the relaxation time is fitted to produce the exact area of $F_s(k,t)$, 
i.e. $S_s(k,\omega=0)$, for large $k$, whereas for low $k$ the correct area is
given in terms of the diffusion coefficient, which does not appear in the 
parametrization used. 
It can also be noticed that $F_{s, \rm visc}(k,t)$ does not vary
monotonously with time for large $k$, showing oscillations which do not 
appear in the MD or in the self-consistent results.
The GA model leads to $F_{s,g}(k, t)$ which compare 
favorably with the MD results, although for the intermediate $k$-values 
predicts a quicker decrease. Taking $z=0$ in Eq. (\ref{MfFs}) 
shows that this is a consequence of the previously mentioned underestimation 
of the corresponding $\Gamma_{s,g}(k, t)$  at long times. 
%%****************************************
%\begin{figure}
%\begin{center}
%\mbox{\psfig{file=f9.eps,angle=-90,width=85mm}}
%\end{center}
%\caption{Same as the previous figure but for T = 725 K. } 
%\label{FsKT725}
%\end{figure}
%%*******************************************************

By Fourier transforming $F_s(k, t)$ we obtain the  
spectrum $S_s(k, \omega)$ which, for all $k$-values, exhibits a 
monotonic decay with frequency, from a peak value at $\omega = 0$. 
In fact, the relevant features embodied in 
 $S_s(k, \omega)$ are conveniently expressed in terms of   
the peak value $S_s(k, \omega = 0)$, and the half-width at 
half maximum, $\omega_{1/2}(k)$. These magnitudes are usually reported %as 
normalized with respect to the values of the diffusive 
($k$ $\to$ 0 ) limit, %namely $(\pi D k^2)^{-1}$ and $Dk^2$ respectively; 
%this is carried out by 
introducing the dimensionless quantities 
$\Sigma(k)=\pi D k^2 S_s(k, \omega =0)$ and 
$\Delta(k) = \omega_{1/2}(k)/D k^2$. The magnitude
$\omega_{1/2}(k)/ k^2$ can be also interpreted as an
effective $k$-dependent diffusion coefficient $D(k)$. 
%It is known that 
For a liquid near the triple point, $\Delta(k)$ usually exhibits an 
oscillatory behaviour whereas, in a dense gas it decreases monotonically 
from unity at $k$ = 0 to the $1/k$ behaviour at large $k$.  The 
results obtained for $\Delta(k)$ in this work are shown in Fig. \ref{Delta}.   
%First, in figure \ref{Delta} 
%we have plotted the obtained 
%results for $\Delta(k)$, as given by different approaches. 
The MD values, $\Delta_{\rm MD}(k)$, show that for all 
temperatures, the diffusive limit
is reached from below, with a minimum at around $k \approx k_p$, followed 
by a maximum and by a gradual transition, for greater $k$-values, 
%the $\Delta_{\rm MD}(k)$ evolves towards merging 
to the free-particle limit.
This oscillating behaviour of $\Delta_{\rm MD}(k)$ for
small and intermediate $k$-values has already been studied by several
authors \cite{Balubook,MorGla,WahnSjo,Verk1,Verk2,Montfrooy} and has
been attributed to the coupling of the single-particle motion
to other modes in the system; in terms of the present theoretical formalism
this effect would be described by the $\Gamma_{s, \rm MC}(k, t)$ term in
Eq. (\ref{Gammastot}).
We note that similar features to those obtained in this paper
for $\Delta_{\rm MD}(k)$, were already obtained by
Torcini {\it et al.} \cite{TorBalVer} in their MD study of liquid 
lithium using the interatomic pair potentials proposed by    
Price {\it et al.} \cite{PST}  

%%***********************************************
%\begin{figure}
%\begin{center}
%\mbox{\psfig{file=f10.eps,angle=-90,width=85mm}}
%\end{center}
%\caption{Normalized half width of $S_s(k, \omega)$, relative to 
%its value at the hydrodynamic limit, for liquid lithium at 
%four temperatures. Open circles: MD results. 
%Continuous line: present theory. Dashed line: GA model for 
%$F_s(k, t)$ . Dotted line: GA model with non-gaussian 
%corrections for  $F_s(k, t)$. Long dashed line: 
%Free particle limit. Dash-dotted line: present 
%theory with only the binary term in Eq. (\ref{Gammastot})}
%\label{Delta}
%\end{figure}
%%**************************************************************
The self-consistent theoretical results obtained for 
$\Delta(k)$ exhibit a qualitative 
agreement with the corresponding MD ones. In particular the  
positions of the minimum and maximum of 
 $\Delta_{\rm MD} (k)$, are successfully reproduced, although 
they are somewhat overeshot. These features 
are closely related to the term $\Gamma_{s,\rm MC}(k, t)$, which gives the 
intermediate and long time contributions of 
the second order memory function.  This is shown by the fact that 
when only the binary term,
$\Gamma_{s,B}(k, t)$, is considered in Eq. (\ref{Gammastot}),
the resulting values for the diffusion coefficient are poor, 
and the corresponding $\Delta_B(k)$ initially increases
with $k$, contrary to the behaviour of the 
MD simulations.\cite{Balubook,WahnSjo} Therefore, 
the inclusion of a tail in  $\Gamma_{s}(k, t)$ seems important  in order to obtain 
the oscillating structure of  $\Delta_{\rm MD} (k)$.

%%***************************************************
%\begin{figure}
%\begin{center}
%\mbox{\psfig{file=f11.eps,angle=-90,width=85mm}}
%\end{center}
%\caption{Normalized peak value $S_s(k, \omega=0)$, relative to 
%its value at the hydrodynamic limit, for liquid lithium at 
%four temperatures. Open circles: MD results. 
%Continuous line: present theory. Dashed line: GA model for the 
%$F_s(k, t)$ . Dotted line: GA model with non-gaussian 
%corrections for  $F_s(k, t)$. Long dashed line: 
%Free particle limit. Dash-dotted line: present 
%theory with only the binary term in Eq. (\ref{Gammastot})}
%\label{Sigma}
%\end{figure}
%%**************************************************************

The self-consistently calculated values of $\Sigma(k)$ 
are plotted in Fig. \ref{Sigma}, 
where the comparison with the MD results shows that the present 
formalism provides a rather good description of this magnitude.   
The shape of $\Sigma(k)$ is more sensitive to 
changes in temperature, with the diffusive limit reached from below for 
T = 470 K and from above for T=725 K and 843 K, with the transition 
being located somewhere around T= 574 K. 

We point out that the present theoretical results for 
$\Delta(k)$ and $\Sigma(k)$ are, to our knowledge, the first that have
been  obtained within a memory function/mode-coupling formalism, 
with no recourse to parameters, or inclusion of MD data, 
at any stage of the calculation.

For comparison, we have also included in Figs. \ref{Delta} and \ref{Sigma} the  
results obtained using the GA model of Eq. (\ref{Fsgauss}) for $F_s(k, t)$, 
using the MD values for $\delta r^2(t)$. This approximation 
is exact for both small and large $k$-values, whereas for intermediate 
$k$-values the decay of  $F_s(k, t)$ is too fast, producing too large a width 
and a smaller initial value of the spectrum 
$S_s(k, \omega)$. This is clearly observed in 
Figs. \ref{Delta} and \ref{Sigma}. In fact, the   
associated $\Delta_{g}(k)$ is unable to 
follow the $k$-dependent behaviour 
exhibited by the corresponding MD results, especially the minimum  
appearing at $k \approx k_p$. Note that the largest 
discrepancies appear for the intermediate $k$-values, for which the 
spatial correlations are stronger, and that the  
discrepancies become smaller as the density is reduced. 
On the other hand, the GA results ontained for $\Sigma_{g}(k)$ qualitatively 
reproduce the MD results, although underestimating them. 
Moreover, as the temperature increases the agreement is improved. 
It is interesting to mention that although the 
$\Delta_{g}(k)$ does not exhibit the oscillatory behaviour of 
 $\Delta_{\rm MD}(k)$, however the corresponding second-order memory 
function, $\Gamma_{s, g}(k, t)$ does have a tail which fairly follows 
the intermediate and long time behaviour displayed by
$\Gamma_{s,\rm MD}(k, t)$, as is observed in 
Figs. \ref{MEMFsKT470} and \ref{MEMFsKT725}. 
Therefore, the existence of a tail in the 
second order memory function of  $F_s(k, t)$, although necessary, does not 
automatically imply the 
appareance of an oscillatory behaviour of  $\Delta(k)$. 

To gain further insight into the role played by the tail of  
 $\Gamma_s(k, t)$ in the 
oscillatory behaviour of  $\Delta (k)$, we have also evaluated 
$F_s(k, t)$ using an extension of the GA model which includes some 
non-gaussian corrections, obtained 
by means of a cumulant expansion.\cite{NijRah} Restricting 
up to the first non-gaussian 
term, which provides the dominant corrections to the gaussian result, 
we have the following expression

\begin{eqnarray}
\label{Fsnongauss}
& & F_s(k, t) = \nonumber \\
& & \left[ 1 + \frac{1}{2} \alpha_2(t)  
\left[ \;  \frac{1}{6} \; k^2 \; \delta r^2(t) \right]^2 \right] \; \; 
\exp \left[ \; -   \frac{1}{6} \; k^2 \; \delta r^2(t) \right]  \;
\end{eqnarray}

\noindent where $\alpha_2(t)$ is the first non-gaussian coefficient:

\begin{equation}
\alpha_2(t) = \frac{3}{5} \; \frac{ \delta r^4(t)}
{ [\delta r^2(t)]^2} \; - \; 1 
\end{equation}

\noindent and  $\delta r^4(t) \equiv 
\langle \mid  \vec{r}_1(t) - \vec{r}_1(0) \mid^4 \rangle$.  
By using as input data the $\delta r^2(t)$ and $\delta r^4(t)$ 
obtained from the MD simulations, we have evaluated the corresponding $F_s(k, t)$, according 
to Eq. (\ref{Fsnongauss}). The associatted $\Delta_{\rm n-g}(k)$ and 
$\Sigma_{\rm n-g}(k)$ accurately reproduce the corresponding MD 
results, as shown in  Figs. \ref{Delta} and \ref{Sigma}. Moreover, the 
corresponding second-order memory functions, $\Gamma_{s, \rm n-g}(k, t)$, 
reproduce the MD results quite accurately. 
This is shown in Fig. \ref{2MEMFsKT470} for T = 470 K, 
where we have focused on the 
intermediate and long time behaviour, and the comparison is also 
performed with the MD simulations, the GA model and the present 
theoretical framework. Note that whereas  
the GA model always underestimates the MD results, the opposite happens 
with the present theoretical framework. The underestimation induced by
the GA model is more marked for those $k$-values around the main 
peak of $S(k)$ which is precisely the region where $\Delta_{\rm MD}(k)$ 
exhibits a minimum. 
In order to explain the conection between this
behaviour and the shape of the corresponding $\Delta(k) \equiv D(k)/D$, we
must note that
$D(k) = \tilde{D}(k, z=0)$, where
 
\begin{equation}
k ^2 \; \tilde{D}(k, z) =  \tilde{K}_s(k, z) \; =
\frac{\Omega_s^2(k)}{z +\tilde{\Gamma}_s(k, z)} \, ,
\end{equation}
 
\noindent where
$\tilde{D}(k, z)$ stands for a generalized
diffussion coefficient.\cite{Balubook} When this equation is
taken at $z=0$ we obtain
that $D(k)\; \tilde{\Gamma}_s(k, z=0) = (1/\beta m)$.
Therefore, according to
Fig. \ref{2MEMFsKT470}, the GA model
leads to a second order memory function, $\Gamma_{s, g}(k, t)$ which
underestimates the corresponding MD results for all $k$-values,
and by the previous relationship it leads to greater estimates of
the corresponding $D(k)$, when compared
with the corresponding MD results. By contrast, the opposite behaviour
is exhibited by the present theoretical formalism as it overestimates
the corresponding $\Gamma_s(k, t)$ and therefore leads to smaller values
of the corresponding $D(k)$.

%%****************************************************
%\begin{figure}
%\begin{center}
%\mbox{\psfig{file=f12.eps,angle=-90,width=85mm}}
%\end{center}
%\caption{Second-order memory function, $\Gamma_s (k, t)$, of 
%the self intermediate scattering functions, $F_s(k, t)$, at  
%several $k$-values, for liquid lithium at T = 470 K. Open circles: MD results. 
%Continuous line: present theory. 
%Dash-dotted line: GA model for  $F_s(k, t)$.  
%Dotted line: GA model with non-gaussian corrections.} 
%\label{2MEMFsKT470}
%\end{figure}
%%*******************************************************************

\subsection{Transport properties}

The normalized velocity autocorrelation functions obtained from
the self-consistently calculated results, using  
Eq. (\ref{VACF}), are shown in 
Fig. \ref{VACF3}, where they a compared with the corresponding MD  results. 
The initial decay of $Z(t)$ is very well reproduced, because its initial 
value and both the second and fourth derivatives are implicitly imposed 
through Eq. (\ref{gammasB}). The positions of the maxima and minima of 
$Z(t)$ are also well predicted, although the amplitude of the oscillations is 
underestimated. The corresponding self-diffusion 
coefficients, $D$, are shown 
in Table \ref{diffu} along with those obtained by the MD simulations. 
%%$$$$$$$$$$$$$$$$$$$$$$$$$$$$$$$$$$$$$$$$$$
%\begin{figure}
%\begin{center}
%\mbox{\psfig{file=f13.eps,angle=-90,width=65mm}}
%\end{center}
%\caption{Normalized velocity autocorrelation functions of liquid Li at 
%three temperatures. Open circles: MD results. Continuous lines: 
%self-consistent calculations.  } 
%\label{VACF3}
%\end{figure}
%%$$$$$$$$$$$$$$$$$$$$$$$$$$$$$$$$$$$$$$$$$$$$$$$$$$$$$$$$$$$$$$$$
For all the temperatures considered the 
MD results agree rather well with the experimental  
INS \cite{Jong1,Seldmeier} and  
tracer data \cite{Lowen} (for 
T = 843 K the experimental data have been extrapolated slightly 
outside the range suggested in Refs. \onlinecite{Jong1} and \onlinecite{Seldmeier}). 
This good agreement supports the adequacy of the NPA-derived 
interatomic pair potentials to describe liquid lithium in this temperature range.
The theoretical results show good agreement with the MD values for  
T=470 K, whereas as the temperature is increased 
the MD results are underestimated. 
In fact, according to the relation between $D$ and 
the VACF, see Eq. (\ref{DifZ}), the smaller values obtained for $D$ are a   
consequence of the above mentioned underestimation for the amplitude of the  
oscillations in  $Z(t)$, especially that of the first maximum. This is more 
evident at T=725 K, whereas for 
T = 470 K there is a cancellation between the first minimum and the 
first maximum leading to a theoretical $D$ very close to the 
MD result.

As regards the shear viscosity, fig. \ref{shearTOT} shows, 
for T=470 and 725 K, the MD results obtained 
for the SACF along with its three contributions. Note that   
$\eta_{kp}(t)$ and $\eta_{kk}(t)$ parts of the SACF are 
very small (less than $10\%$ of the potential-potential part) 
which justifies their being ignored in the theoretical calculations of the 
SACF. 
The figure also shows the theoretical $\eta(t)$ along with 
its binary and mode-coupling components. Note that whereas for the 
lower temperature the inclusion of the mode-coupling component 
is essential for achieving a good overall agreement with the 
MD data, when the temperature is increased the role of the binary part 
becomes more dominant; in fact, for 725 K   
the binary part alone accounts for most of the MD results.

Our present theoretical results  
slightly overestimate the short time ($t \approx 0.1$ ps) 
behaviour of $\eta(t)$ and 
this is mainly due to the binary component; 
more explicitly, it comes from the overestimation of the 
values for  $\tau_{\eta}$. This limitation can be traced back to the
use of the superposition approximation in the evaluation of the
three-body term appearing in $\tau_{\eta}$, which has a rather
important weight in the final value.\cite{Alemany}

On the other hand, the behaviour for $ t > 0.1$,  
which is completely determined by the mode-coupling component, 
is rather well described for both temperatures. 
The values obtained for the shear viscosity coefficient, $\eta$, are 
presented in Table \ref{visco}.  
Although the theoretical values 
slightly overestimate 
the MD ones (because of the short time behaviour of $\eta(t)$),
both the theoretical and
the MD results 
agree well with 
the experimental values.\cite{Shpil} 

%%*****************************************************************
%\begin{figure}
%\begin{center}
%\mbox{\psfig{file=f14.eps,angle=-90,width=85mm}}
%\end{center}
%\caption{Normalized potential part of the stress 
%autocorrelation function, $\eta(t)$,  
%for liquid lithium at T=470 and 725 K. Open 
%circles: MD results. Continuous line: theoretical results.
%Dash-dotted line: binary part. Dotted line: mode-coupling component. 
%The inset shows the MD results for $\eta(t)$ 
%(open circles), $\eta_{pp}(t)$ (continuous line), 
% $10 \times \eta_{kp}(t)$ (dotted line) 
%and $10 \times \eta_{kk}(t)$ (dashed line). } 
%\label{shearTOT}
%\end{figure}
%%*******************************************************************

\section{Conclusions}

In this work we have evaluated several 
dynamical properties of liquid lithium at thermodynamic 
states close to the triple point. The calculations were performed by a 
self-consistent theoretical framework which, by  
incorporating mode-coupling concepts, allows the  
evaluation of both  single-particle properties, as represented by the 
self-intermediate scattering function,   
its memory functions, the velocity autocorrelation function 
and the self-diffusion coefficient, 
as well as collective properties such as 
the intermediate scattering function, its memory functions, 
the autocorrelation function of the 
non-diagonal elements of the stress tensor, and the shear viscosity 
coefficient. Its application to  
liquid lithium has led to reasonable results when compared with both  
the corresponding MD results and the available experimental data. 
The agreement 
is particularly satisfactory near the melting point, deteriorating  somewhat 
at higher temperatures.  

Within the present theoretical formalism, 
memory functions play a key role, specifically 
the second order memory 
function of both the intermediate scattering function and its self counterpart, 
as defined in Eq. (\ref{MfF}) and (\ref{MfFs}).
All the relaxation 
mechanisms controlling the collective and single-particle dynamics, are 
introduced at the level of the corresponding second order memory 
functions which are decomposed
into their binary and mode-coupling contributions. 
We have found that the 
gaussian ansatz adopted for the binary term 
provides a reasonable description of the corresponding MD results
for short times.
As for the other contribution, namely the mode coupling part, for simplicity we 
have only considered the density-density coupling term. 
Although other couplings could be included, this term 
provides the dominant mode coupling contribution  
for thermodynamic states close to the triple point, which explains 
why both $\Gamma(k, t)$ and $\Gamma_s(k, t)$ are 
better described for the lower temperatures.  
 
The self-consistent theoretical calculations provide results for 
$\Gamma(k,t)$ and $F(k,t)$ in qualitative agreement with the MD
results, and much more accurate than
%In the case of  $\Gamma(k, t)$, the present framework qualitatively accounts 
%for the corresponding MD results and the same happens for the associatted 
%$F(k, t)$. Despite of several shortcomings, this approach represents 
%an improvement when compared with 
other theoretical approaches, as 
for example the widely used viscoelastic model. 
%Moreover, we stress that 
%the evaluation of  $F(k, t)$ is directly performed from its 
%memory function/mode coupling formalism. 

Better results have  been achieved for 
 $\Gamma_s(k, t)$ and the corresponding 
$F_s(k, t)$. Within this context, we emphasize the good description 
%provided by the present theoretical results for the 
of the wavevector dependence of both the  
peak height and the half-width at half maximum of the 
$S_s(k, \omega)$, as represented by $\Sigma(k)$ and $\Delta(k)$ 
respectively. Since these funcitons constitute a stringent test 
of any theoretical model,\cite{Balubook} we conclude that 
the main physical effects behind  $F_s(k, t)$ seem to be included in 
the present theoretical framework, specifically, in the 
expression adopted for $\Gamma_s(k,t)$.
We stress that, to our knowledge, 
this is the first theoretical study where the behaviour of  $\Delta (k)$  
has been qualitatively reproduced from just its memory function/mode-coupling 
formalism, without resorting to parameters, fitting to an assumed 
shape or including magnitudes from MD simulations in the evaluation of 
 $F_s(k, t)$. 

The calculations carried out for the 
$\Gamma_s(k, t)$ have confirmed that the existence of a tail is 
necessary to account for the oscillatory behaviour of 
$\Delta(k)$; however, the magnitude of the tail, understood in the sense 
of its time integral, has a most important 
influence on the amplitude of the oscillations of 
$\Delta(k)$. Moreover, the results obtained for the GA model with 
and without non-gaussian corrections, point towards
the existence of a minimum magnitude of the tail 
in order to induce an oscillatory behaviour on  $\Delta(k)$. 

The improvements achieved in the description of 
 $F_s(k, t)$ and related 
magnitudes, when compared with those predicted by the 
GA model,\cite{CaGG} are not fully reflected in the 
obtained values for the self-diffusion coefficients. However, this is not 
surprising because they are defined as a time integral of a  
correlation function which gives a measure of its time average but 
provides very scarce information on the dynamics 
of the system.

We end up by signaling some limitations of the formalism presented
here. First, its density/temperature range of applicability lies within the
region where the relevant slow relaxation channel is provided by the
coupling to density fluctuations, and this ceases to be valid for densities
smaller than those typical of the melting point. For these
densities, coupling to other modes, like longitudinal and/or transverse
currents, becomes increasingly important and we believe that this 
is the main reason 
for the small deviations observed in the VACF total memory function at the 
two higher temperatures studied. In fact, further 
%Moreover, we consider that an 
improvements in the description of the  $F_s(k, t)$ and $F(k, t)$, 
%intermediate scattering function, $F(k, t)$, 
at all temperatures, would also require 
the inclusion of other modes in the corresponding second-order memory 
function. Moreover, although the previous remarks concern the mode-coupling 
contribution of the second order memory function, attention should also be 
drawn to the binary term. In fact, this term is poorly known and its role 
becomes increasingly dominant with large $k$ and/or at increasing 
temperatures when the mode-coupling contribution decreases. A first task 
would be to study the influence of the superposition approximation for the 
three particle distribution function, on the obtained results for the 
relaxation times $\tau_l(k)$ and $\tau_s(k)$.

Further work is currently performed in that 
direction, and the results will be reported in due time.

\section*{Acknowledgements}

We thank Dr. U. Balucani for helpful comments. 
This work has been supported by the Junta de Castilla 
y Le\'on (Project No. VA70/99), NATO (CRG971173), the British-Spanish 
Joint Research Programme  (HB1997-0188), the
Xunta de Galicia (Project No. PGIDT99PXI20604B), and the CICYT, Spain
(Project No. PB98-0368-C02).

%********************************************
\begin{figure}
%\begin{center}
%\mbox{\psfig{file=f1.eps,angle=-90,width=75mm}}
%\end{center}
\caption{Relaxation times $\tau_l(k)$ and $\tau_s(k)$,  
for liquid lithium. 
Continuous line and dotted line:  $\tau_l(k)$ and $\tau_s(k)$ for T= 470 K. 
Dashed line and dash-dotted line:  $\tau_l(k)$ and $\tau_s(k)$ for T= 725 K.} 
\label{Taus}
\end{figure}
%********************************************

%*************************************************************
\begin{figure}
%\begin{center}
%\mbox{\psfig{file=f2.eps,angle=-90,width=65mm}}
%\end{center}
\caption{Normalized second-order memory function, $\Gamma (k, t)$, of 
the intermediate scattering function, $F(k, t)$, at  
two $k$-values, for liquid lithium at T = 470 K. Open circles: MD results. 
Continuous line: present theory. 
Dash-dotted line: binary part, $\Gamma_B(k, t)$. 
Dotted line: mode-coupling part, $\Gamma_{\rm MC}(k, t)$. 
Dashed line: viscoelastic model. The inset shows the normalized 
first-order memory 
function as obtained by MD (open circles), the viscoelastic 
model (dashed line) and the present theory (continuous line).}
\label{MEMFKT470}
\end{figure}
%**************************************************************

%***********************************************************
\begin{figure}
%\begin{center}
%\mbox{\psfig{file=f3.eps,angle=-90,width=65mm}}
%\end{center}
\caption{Same as the previous figure but for T = 725 K. } 
\label{MEMFKT725}
\end{figure}
%***********************************************************

%**************************************************
\begin{figure}
%\begin{center}
%\mbox{\psfig{file=f4.eps,angle=-90,width=85mm}}
%\end{center}
\caption{Normalized intermediate scattering functions, $F(k, t)$, at  
several $k$-values, for liquid 
lithium at T = 470 K. Open circles: MD results. 
Continuous line: present theory. 
Dashed line: viscoelastic model.}
\label{FKT470}
\end{figure}
%************************************************************

%****************************************************
\begin{figure}
%\begin{center}
%\mbox{\psfig{file=f5.eps,angle=-90,width=85mm}}
%\end{center}
\caption{Same as the previous figure but for T = 725 K. } 
\label{FKT725}
\end{figure}
%******************************************************************

%****************************************************
\begin{figure}
%\begin{center}
%\mbox{\psfig{file=f6.eps,angle=-90,width=85mm}}
%\end{center}
\caption{Normalized second-order memory function, $\Gamma_s (k, t)$, of 
the self intermediate scattering functions, $F_s(k, t)$, at  
several $k$-values, for liquid lithium at T = 470 K. Open circles: MD results. 
Continuous line: present theory. 
Dotted line: mode-coupling part, $\Gamma_{s, \rm MC}(k, t)$. 
Dashed line: viscoelastic model. 
Dash-dotted line: GA model for  $F_s(k, t)$.  
The inset shows the normalized  
first-order memory 
function as obtained by MD (open circles), the viscoelastic 
model (dashed line) and the present theory (continuous line).}
\label{MEMFsKT470}
\end{figure}
%*******************************************************************

%******************************************************
\begin{figure}
%\begin{center}
%\mbox{\psfig{file=f7.eps,angle=-90,width=85mm}}
%\end{center}
\caption{Same as the previous figure but for T = 725 K. } 
\label{MEMFsKT725}
\end{figure}
%**************************************************************

%****************************************
\begin{figure}
%\begin{center}
%\mbox{\psfig{file=f8.eps,angle=-90,width=85mm}}
%\end{center}
\caption{Self intermediate scattering functions, $F_s(k, t)$, at  
several $k$-values, for liquid lithium at T = 470 K. Open circles: MD results. 
Continuous line: present theory. Dashed line: viscoelastic model. 
Dash-dot line: GA model }
\label{FsKT470}
\end{figure}
%*******************************************************

%****************************************
\begin{figure}
%\begin{center}
%\mbox{\psfig{file=f9.eps,angle=-90,width=85mm}}
%\end{center}
\caption{Same as the previous figure but for T = 725 K. } 
\label{FsKT725}
\end{figure}
%*******************************************************

%***********************************************
\begin{figure}
%\begin{center}
%\mbox{\psfig{file=f10.eps,angle=-90,width=85mm}}
%\end{center}
\caption{Normalized half width of $S_s(k, \omega)$, relative to 
its value at the hydrodynamic limit, for liquid lithium at 
four temperatures. Open circles: MD results. 
Continuous line: present theory. Dashed line: GA model for 
$F_s(k, t)$ . Dotted line: GA model with non-gaussian 
corrections for  $F_s(k, t)$. Long dashed line: 
Free particle limit. Dash-dotted line: present 
theory with only the binary term in Eq. (\ref{Gammastot})}
\label{Delta}
\end{figure}
%**************************************************************

%***************************************************
\begin{figure}
%\begin{center}
%\mbox{\psfig{file=f11.eps,angle=-90,width=85mm}}
%\end{center}
\caption{Normalized peak value $S_s(k, \omega=0)$, relative to 
its value at the hydrodynamic limit, for liquid lithium at 
four temperatures. Open circles: MD results. 
Continuous line: present theory. Dashed line: GA model for the 
$F_s(k, t)$ . Dotted line: GA model with non-gaussian 
corrections for  $F_s(k, t)$. Long dashed line: 
Free particle limit. Dash-dotted line: present 
theory with only the binary term in Eq. (\ref{Gammastot})}
\label{Sigma}
\end{figure}
%**************************************************************

%****************************************************
\begin{figure}
%\begin{center}
%\mbox{\psfig{file=f12.eps,angle=-90,width=85mm}}
%\end{center}
\caption{Second-order memory function, $\Gamma_s (k, t)$, of 
the self intermediate scattering functions, $F_s(k, t)$, at  
several $k$-values, for liquid lithium at T = 470 K. Open circles: MD results. 
Continuous line: present theory. 
Dash-dotted line: GA model for  $F_s(k, t)$.  
Dotted line: GA model with non-gaussian corrections.} 
\label{2MEMFsKT470}
\end{figure}
%*******************************************************************

%*******************************************************************
\begin{figure}
%\begin{center}
%\mbox{\psfig{file=f13.eps,angle=-90,width=65mm}}
%\end{center}
\caption{Normalized velocity autocorrelation functions of liquid Li at 
three temperatures. Open circles: MD results. Continuous lines: 
self-consistent calculations.  } 
\label{VACF3}
\end{figure}
%*******************************************************************

%*****************************************************************
\begin{figure}
%\begin{center}
%\mbox{\psfig{file=f14.eps,angle=-90,width=85mm}}
%\end{center}
\caption{Normalized potential part of the stress 
autocorrelation function, $\eta(t)$,  
for liquid lithium at T=470 and 725 K. Open 
circles: MD results. Continuous line: theoretical results.
Dash-dotted line: binary part. Dotted line: mode-coupling component. 
The inset shows the MD results for $\eta(t)$ 
(open circles), $\eta_{pp}(t)$ (continuous line), 
 $10 \times \eta_{kp}(t)$ (dotted line) 
and $10 \times \eta_{kk}(t)$ (dashed line). } 
\label{shearTOT}
\end{figure}
%*******************************************************************

\begin{table}
\caption{Thermodynamic states studied in this work.} 
\label{thermodynamic}
\begin{tabular}{ccccc}
T (K) & 470 & 574 & 725 & 843 \\
\hline
$\rho$ (\AA$^{-3}$) & 0.0445 & 0.0438 & 0.0420 & 0.0416 \\
\end{tabular}
\end{table}

\begin{table}
\caption{Self-diffusion coefficient (in \AA$^2$/ps units),  
of liquid lithium at the thermodynamic states studied in this work. 
$D_{\rm th}$, 
%$D_{\rm visc}$ 
and $D_{\rm MD}$ are the 
theoretical, viscoelastic 
and Molecular Dynamics  
results obtained in this work.}
\label{diffu}
\begin{tabular}{ccccc}
T (K) & 470 & 574 & 725 & 843 \\
\hline
$D_{\rm th}$ & 0.65 & 0.99 & 1.59 & 2.02 \\
%$D_{\rm visc}$ & 0.74 & 1.07 & -- & 2.12 \\  
$D_{\rm MD}$ & 0.69 & 1.11 & 1.94 & 2.47 \\
$D_{\rm exp}$ & 0.64$\pm$0.1 \tablenotemark[1]  
& 1.08$\pm$0.15 \tablenotemark[1] 
& 1.76$\pm$0.25 \tablenotemark[1]   
& 2.28$\pm$0.30 \tablenotemark[1] \\ 
 & 0.69$\pm$0.12 \tablenotemark[2]  
& 1.19$\pm$0.20 \tablenotemark[2] 
& 1.99$\pm$0.30 \tablenotemark[2] 
& 2.62$\pm$0.30 \tablenotemark[2] \\
 & 0.67$\pm$0.06 \tablenotemark[3]  
& 1.16$\pm$0.09 \tablenotemark[3] 
& 1.96$\pm$0.2 \tablenotemark[4] &  \\ 
&    \\ 
% & - & 0.84$\pm$0.06 \tablenotemark[4] & - \\
\end{tabular}
\tablenotetext[1]{Ref.\  \protect\onlinecite{Jong1}}
\tablenotetext[2]{Ref.\  \protect\onlinecite{Seldmeier}}
\tablenotetext[3]{Ref.\  \protect\onlinecite{Lowen}}
\tablenotetext[4]{Ref.\  \protect\onlinecite{Larsson}}
\end{table}

\begin{table}
\caption{Shear viscosity (in GPa ps) of liquid lithium at the  
thermodynamic states studied in this work. $\eta_{\rm th}$ and 
$\eta_{\rm MD}$ are 
the theoretical and Molecular Dynamics results obtained in this work. }
\label{visco}
\begin{tabular}{cccccc}
T (K) & 470 & 574 & 725 & 843 \\
\hline
$\eta_{\rm th}$ & 0.59 & 0.46 & 0.36 & 0.29 \\
%$\eta_{\rm visc}$ & 0.525 & 0.481 & 0.440 \\
$\eta_{\rm MD}$ & 0.55 & 0.42 & 0.33 & 0.28 \\
$\eta_{\rm exp}$ & 0.57$\pm$0.03 \tablenotemark[1] & 
0.45$\pm$0.03 \tablenotemark[1] & 0.35$\pm$0.03 \tablenotemark[1] &  
0.30$\pm$0.03 \tablenotemark[1] \\
\end{tabular}
\tablenotetext[1]{Ref.\  \onlinecite{Shpil}} 
\end{table}

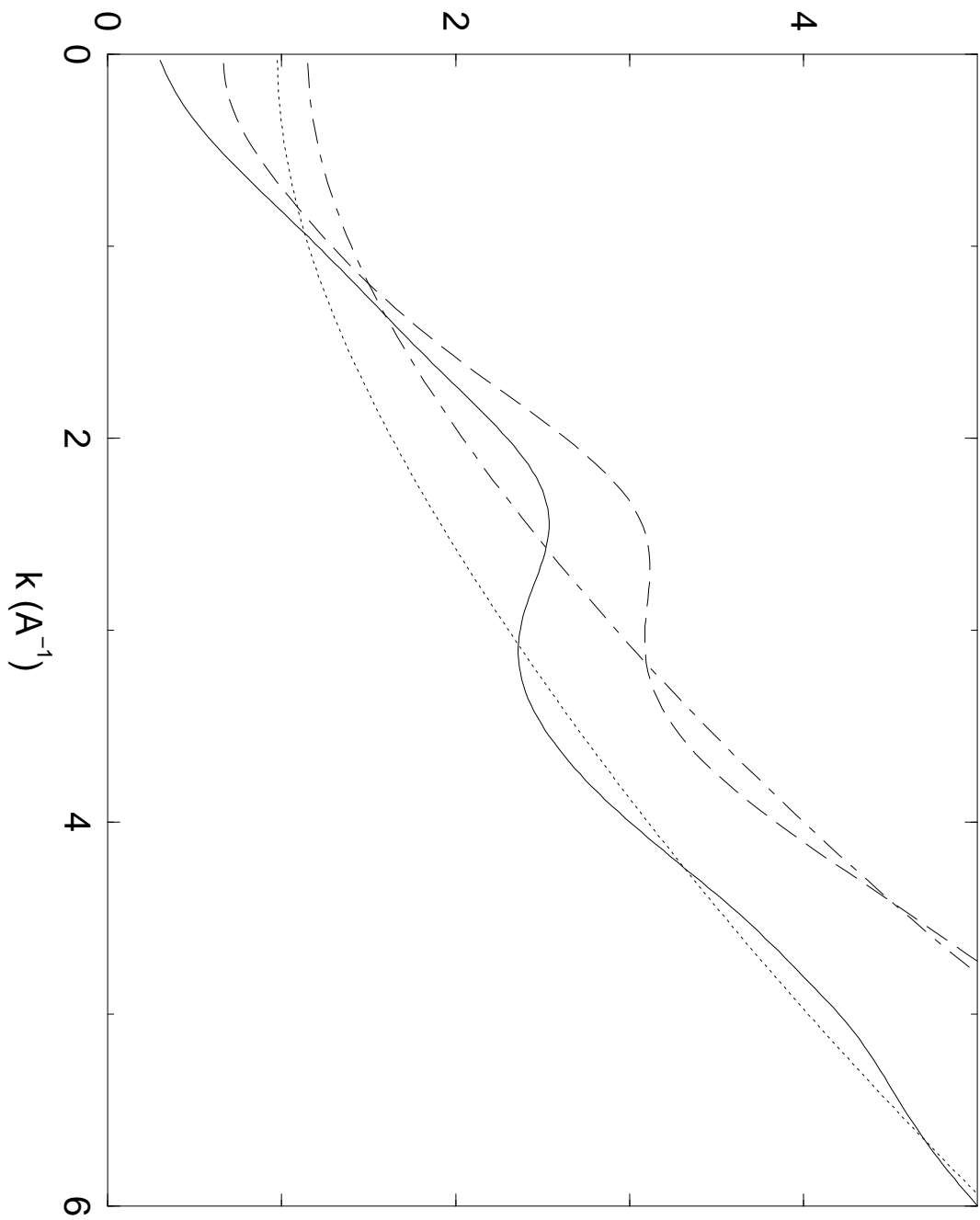
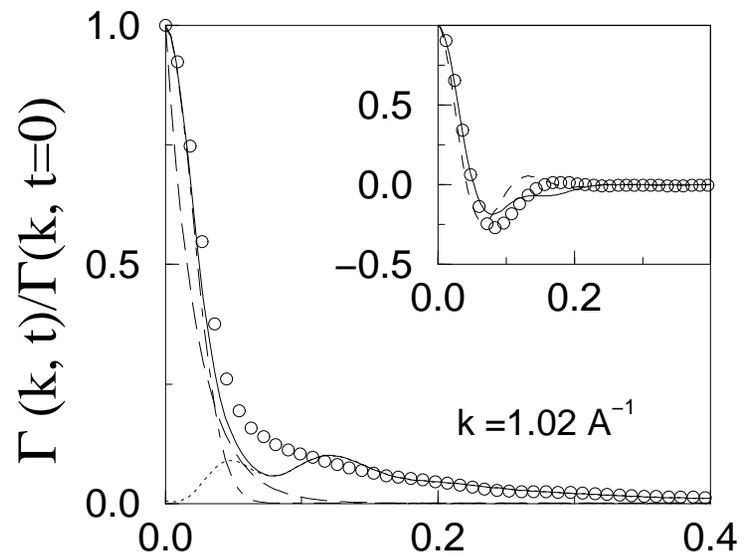
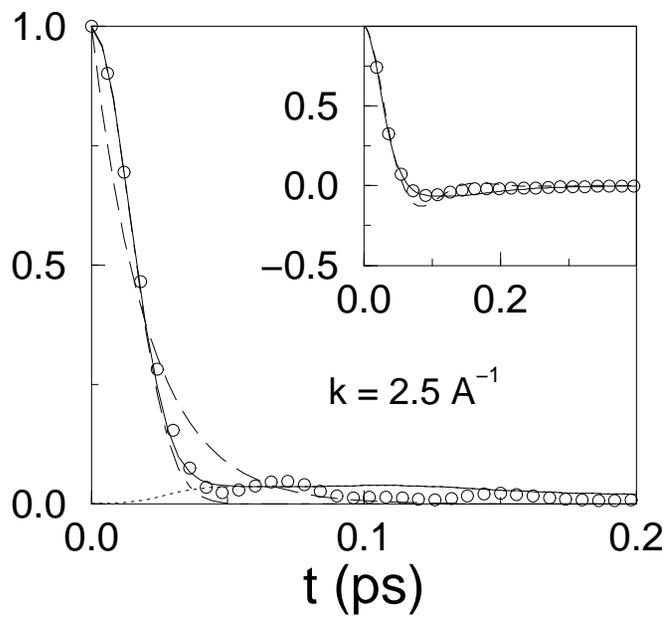
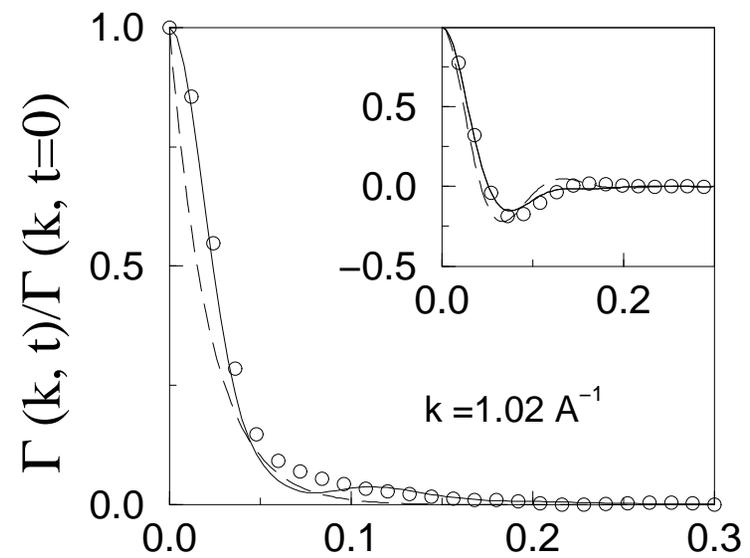
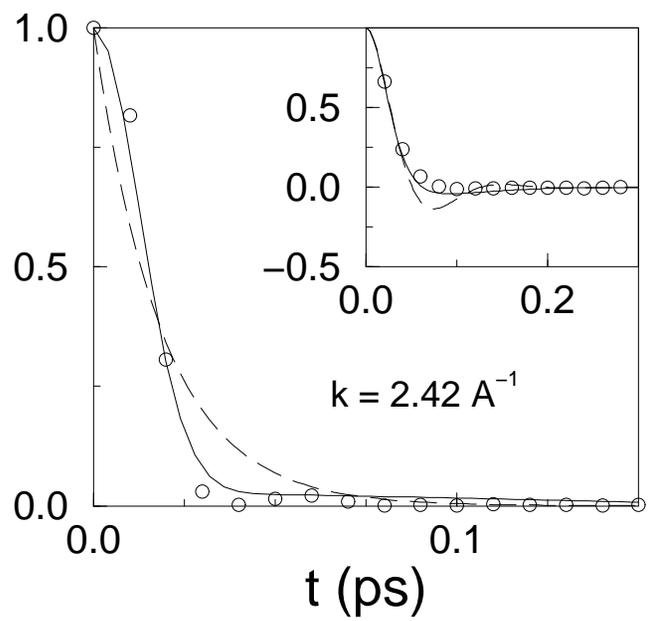
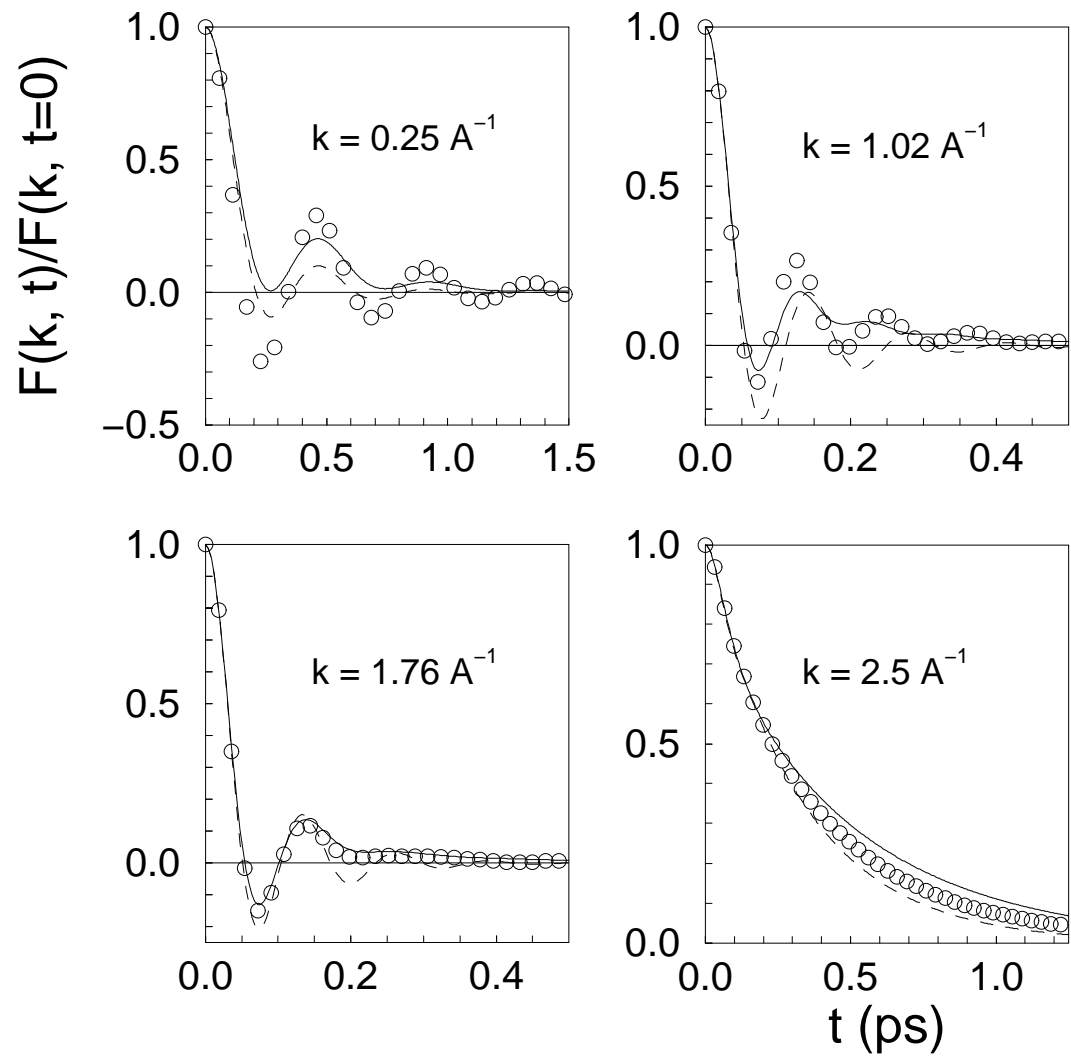
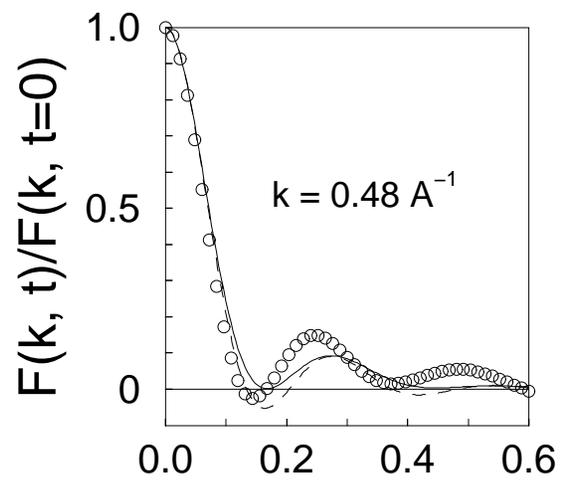
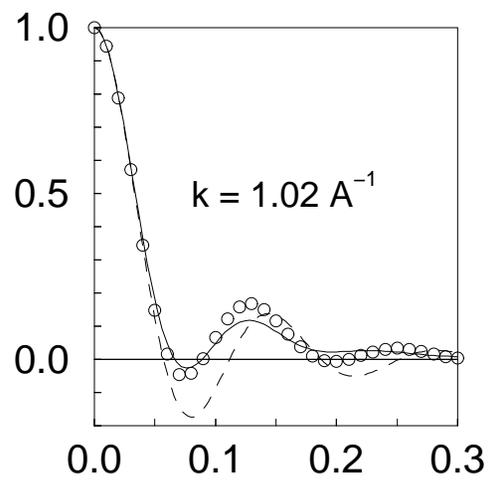
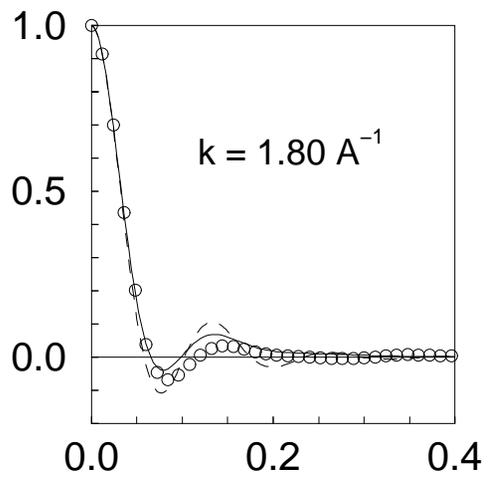
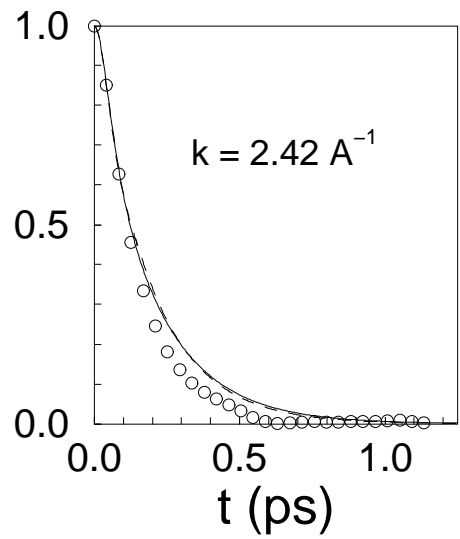
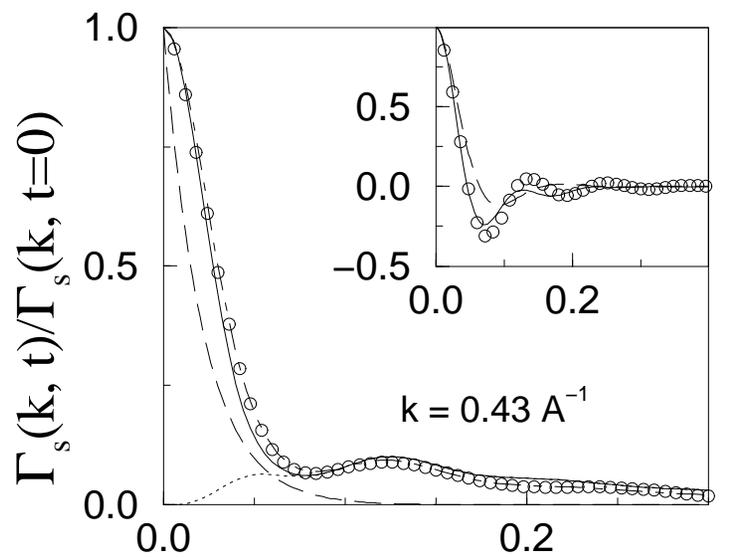
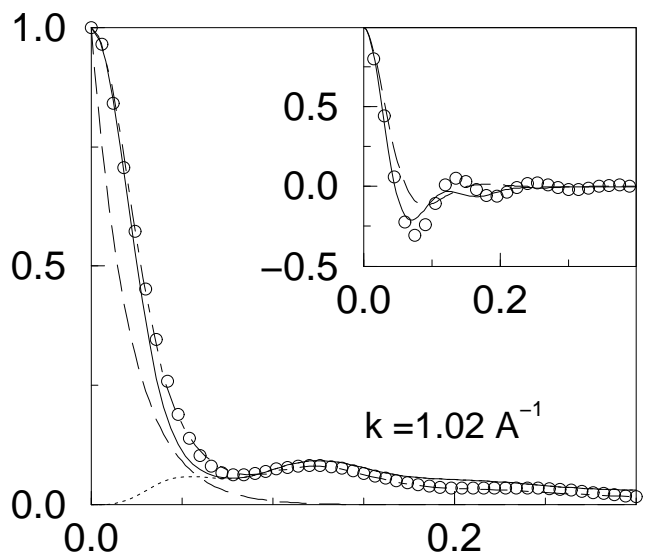
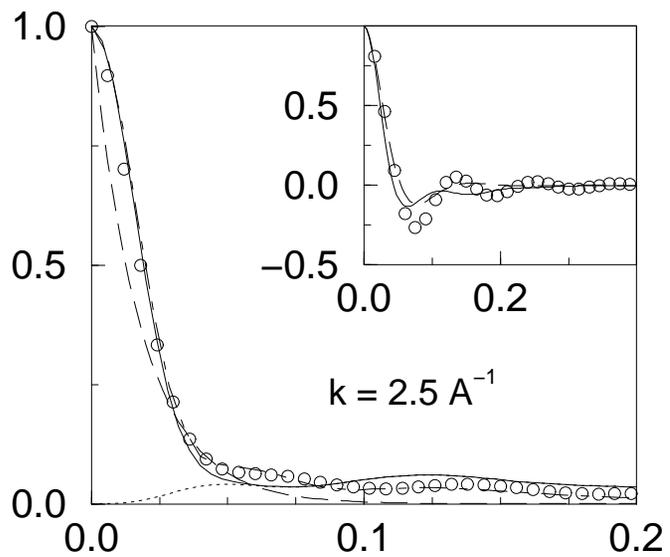
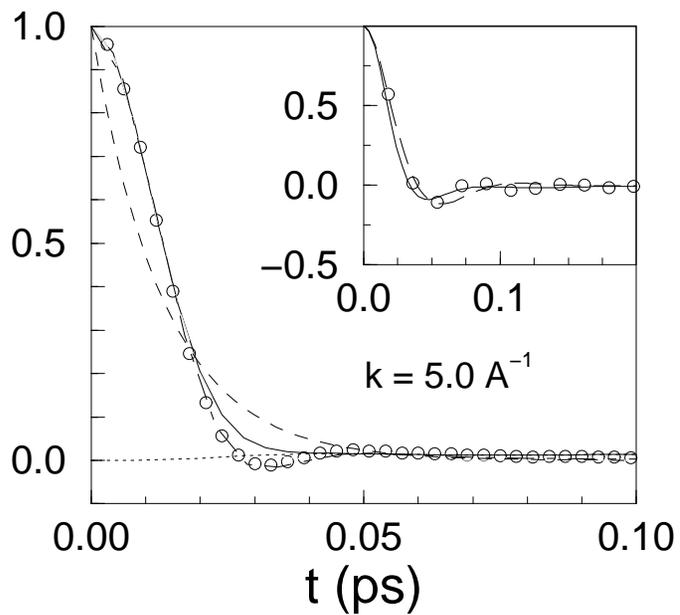
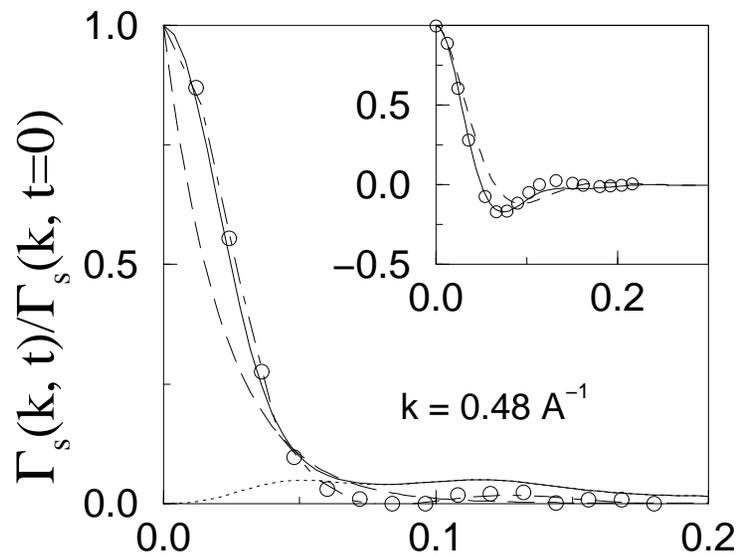
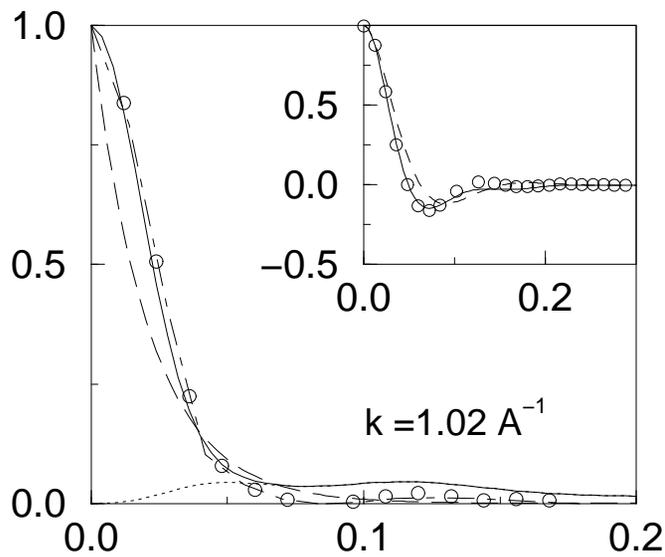
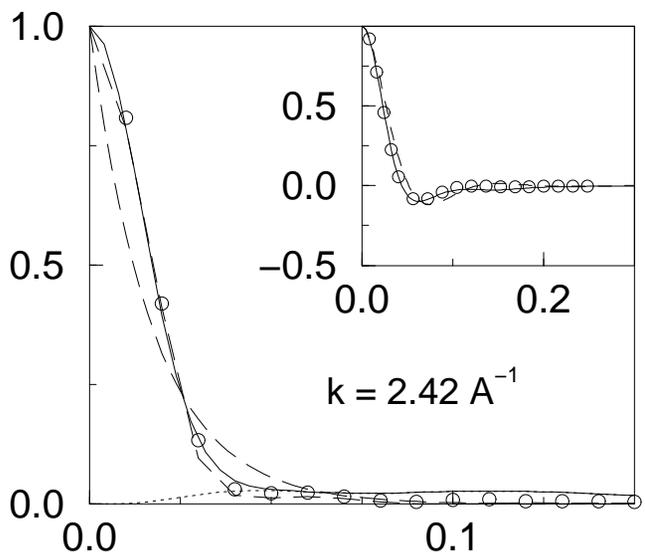
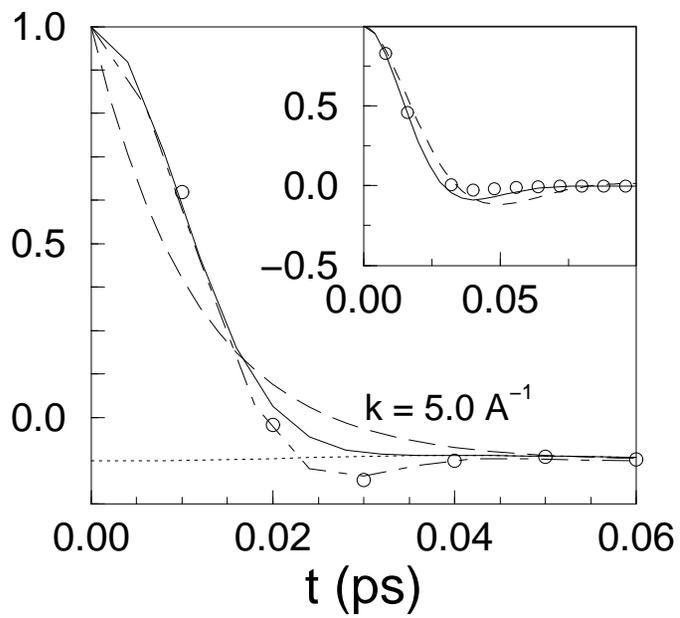
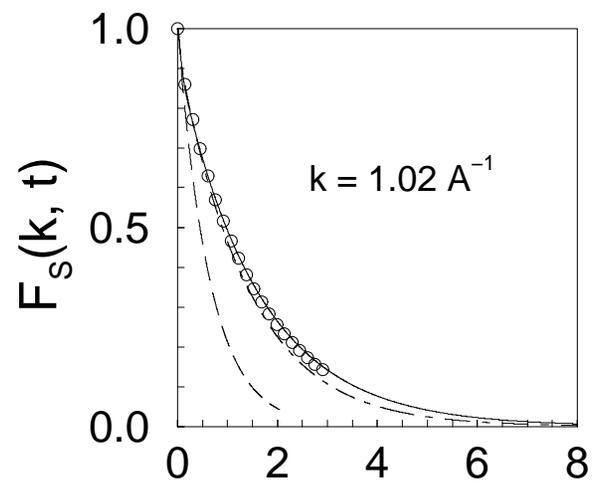
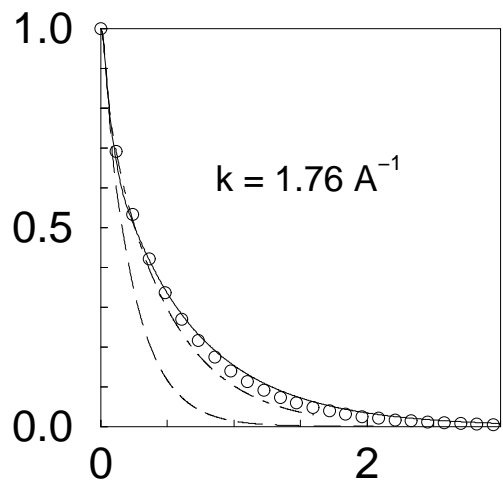
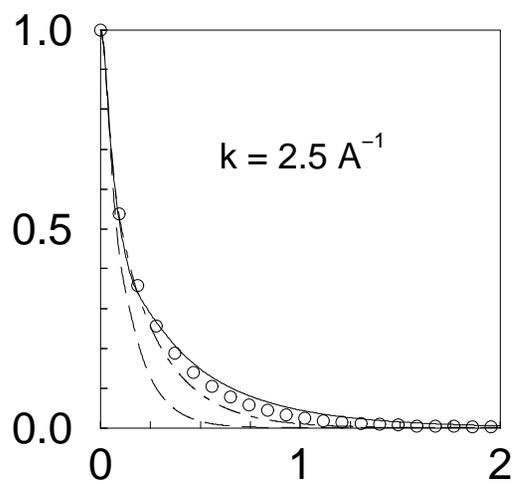
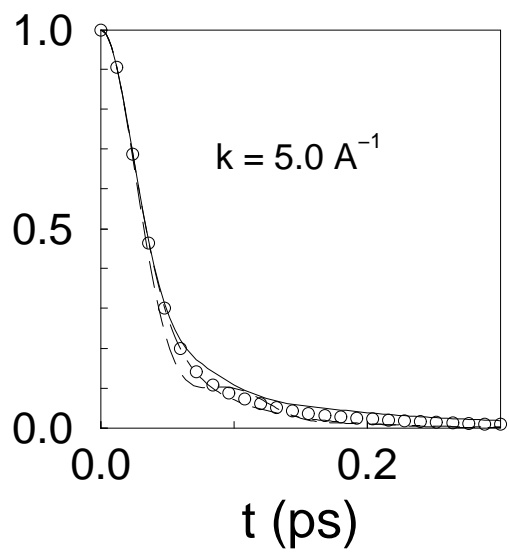
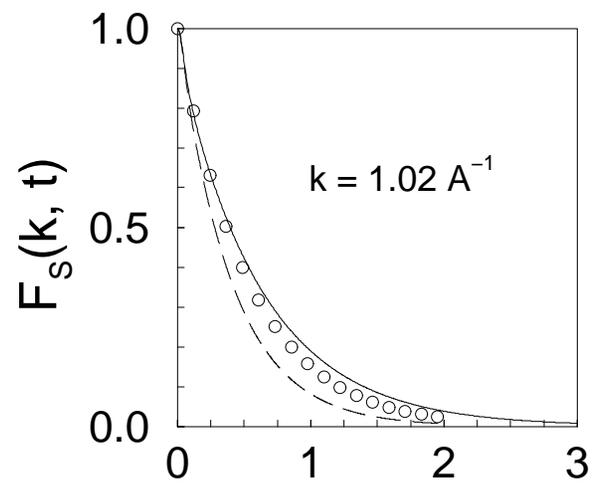
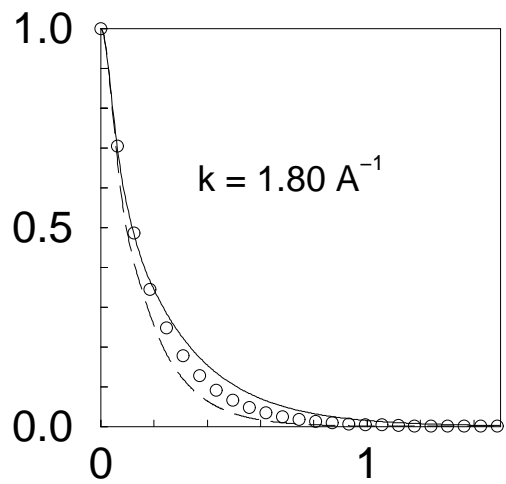
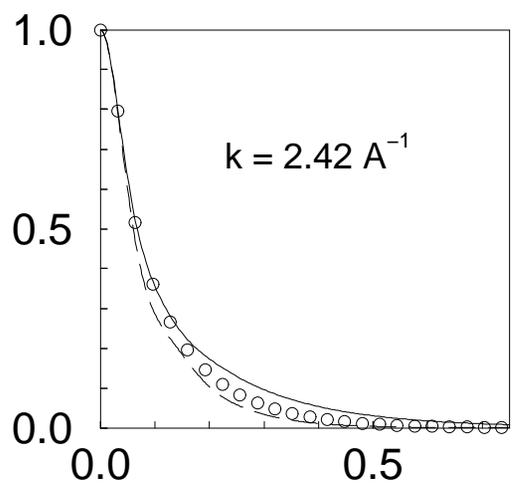
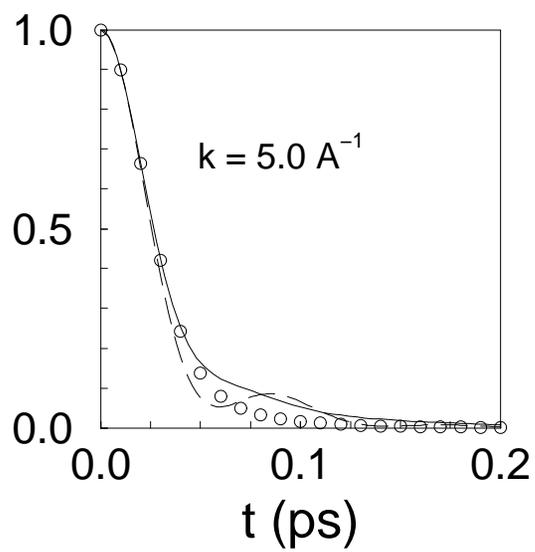
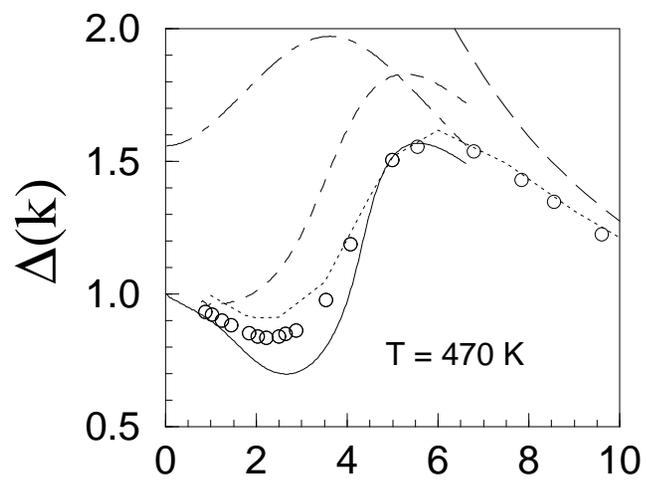
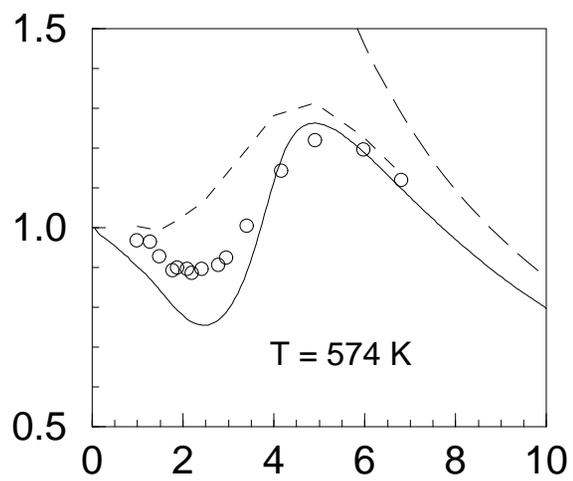
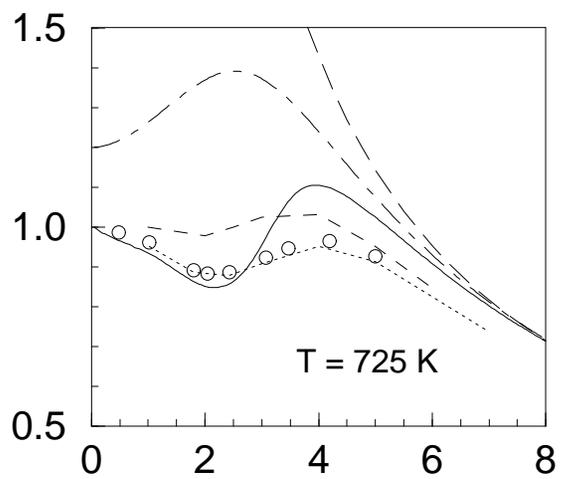
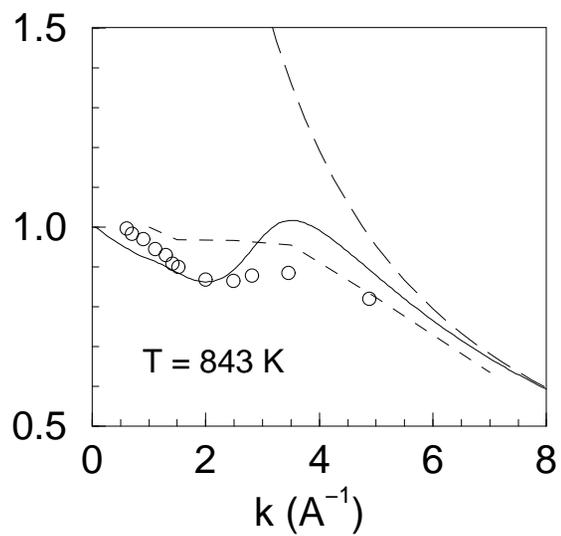
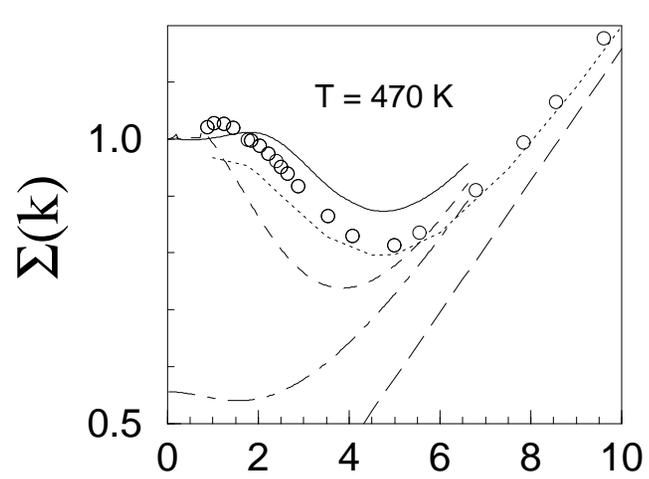
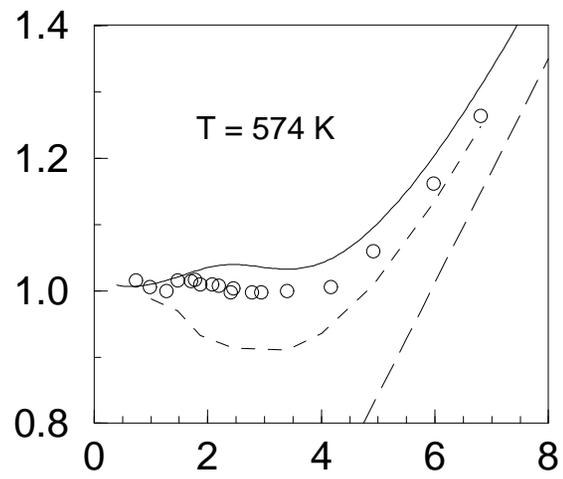
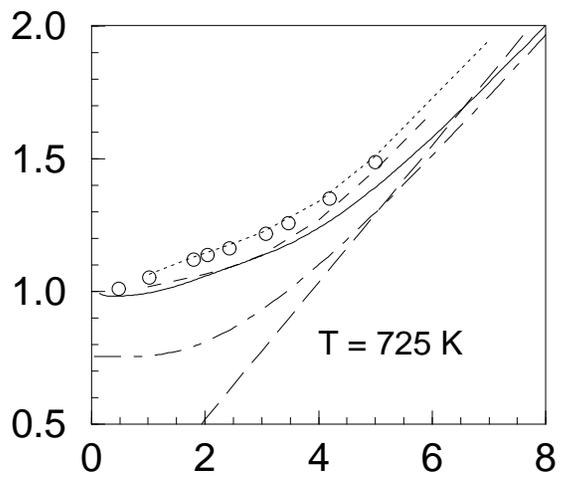
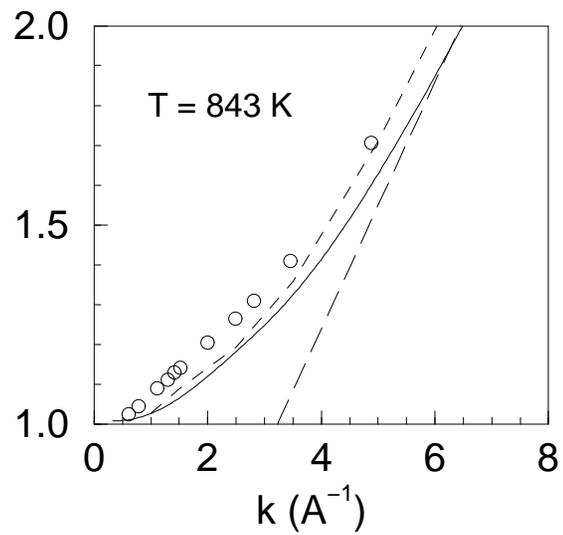
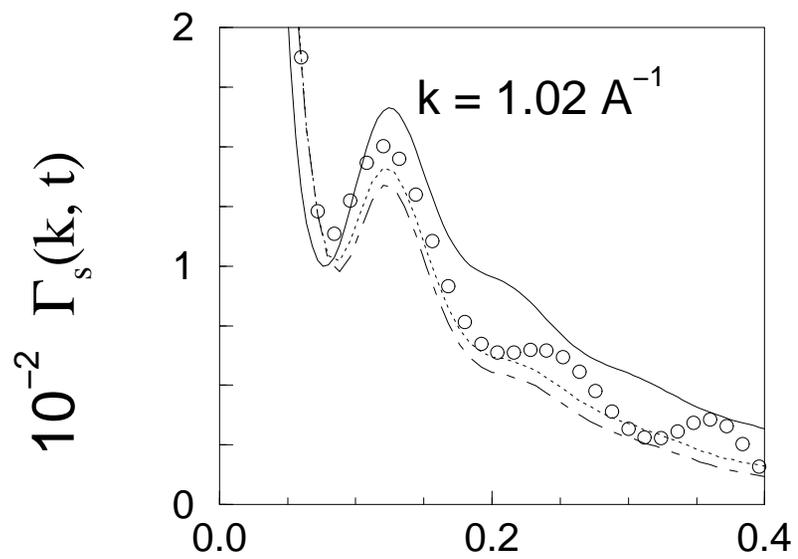
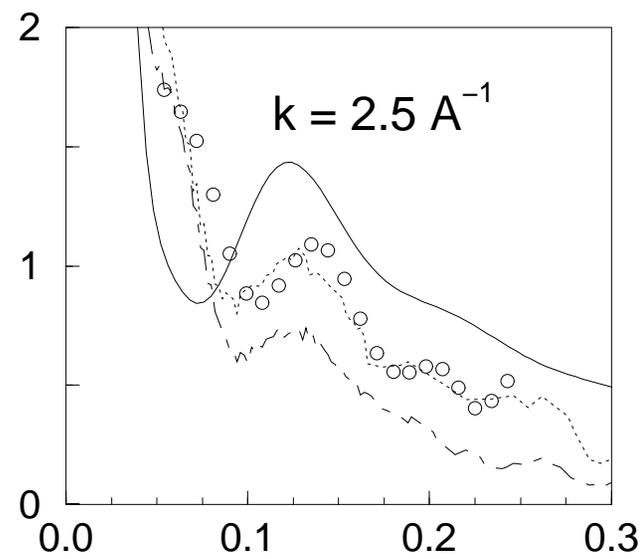
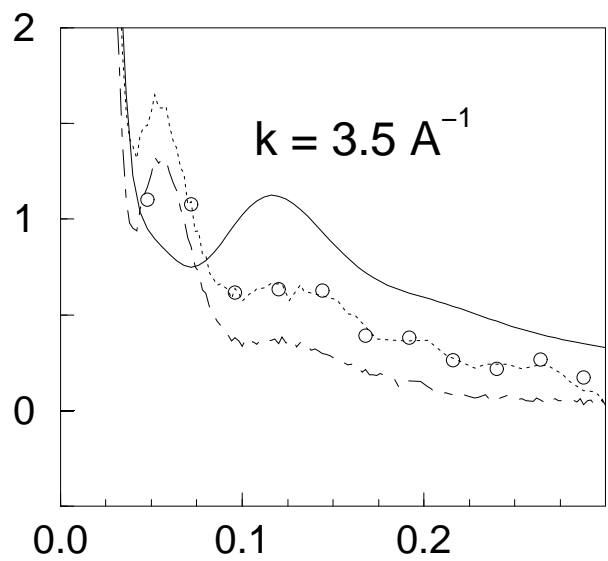
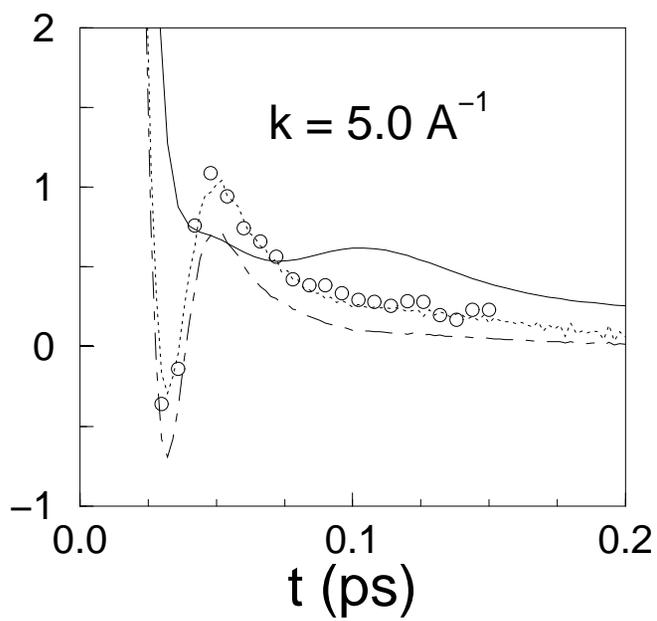
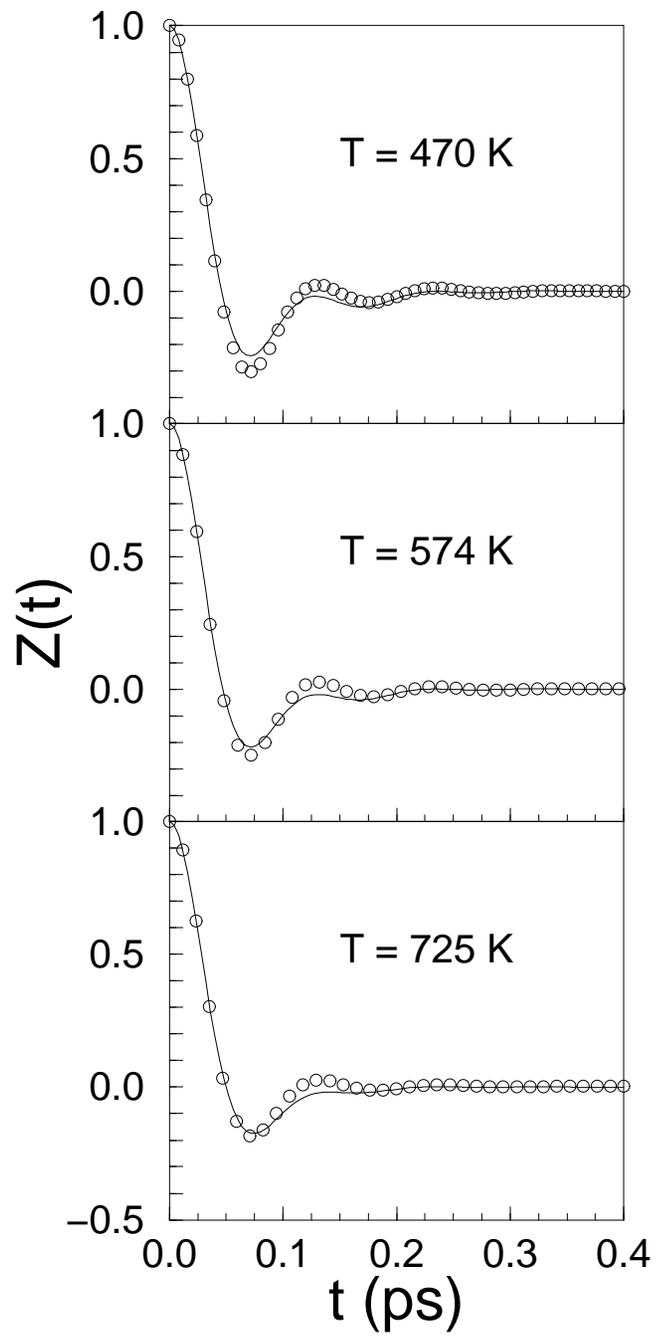
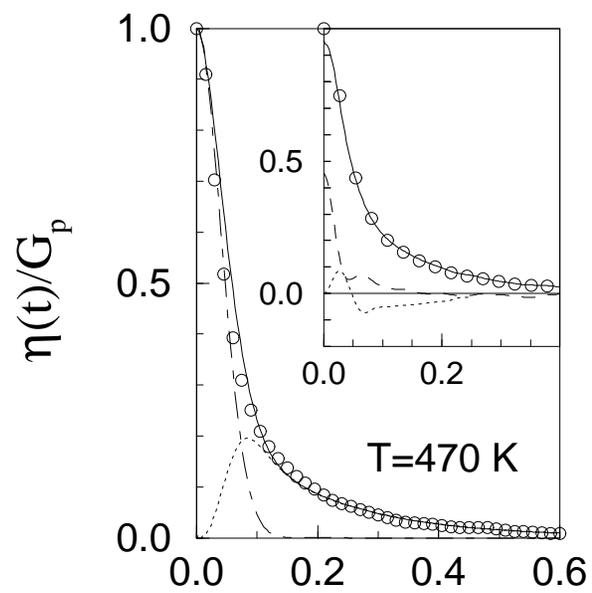
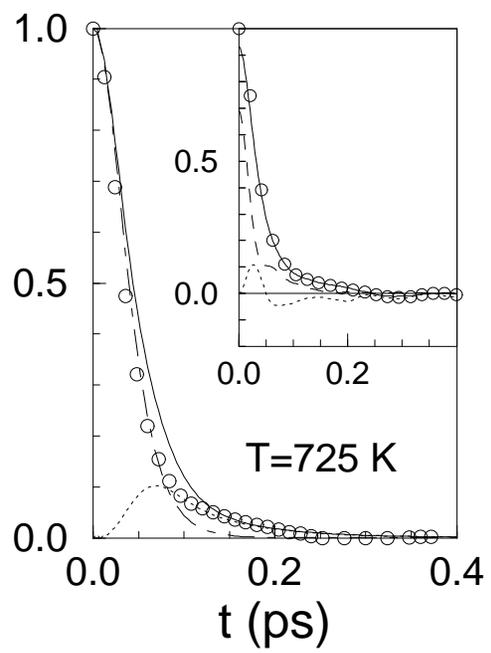
\end{document}